\documentclass[pre,twocolumn,superscriptaddress]{revtex4-1}

\usepackage{amsmath}
\usepackage[xdvi]{graphicx}%
\usepackage{dcolumn}
\usepackage{color}
\usepackage{bm}
\usepackage{subfigure}
\usepackage{array}

\usepackage{amsfonts}

\usepackage[utf8]{inputenc}
\usepackage[T1]{fontenc}
\usepackage{ae,aecompl}
\usepackage{listings}

\usepackage{color}
\definecolor{darkblue}{rgb}{0,0,0.6}
\definecolor{darkred}{rgb}{0.6,0,0}
\definecolor{darkgreen}{rgb}{0,0.6,0}

\usepackage[colorlinks=true,urlcolor=darkblue,citecolor=darkblue,linkcolor=darkred,hyperfootnotes=false]{hyperref}

\newcommand{\bv}[1]{{\boldsymbol #1}}

\newcommand{\vp}{v_{\rm p}}

\newcommand{\Wact}{W_{\rm a}}

\begin{document}

\title{Optimizing active work: 
Dynamical phase transitions, collective motion, and jamming}

\begin{abstract}
Active work measures how far the local self-forcing of active
particles translates into real motion. Using Population Monte Carlo
methods, we investigate large deviations in the active work for
repulsive active Brownian disks.  Minimizing the active work
generically results in dynamical arrest; in contrast, despite the lack
of aligning interactions, trajectories of high active work correspond
to a collectively moving, aligned state. 
We use heuristic and analytic arguments
to explain the origin of dynamical phase transitions separating the
arrested, typical, and aligned regimes.
\end{abstract}

\author{Takahiro Nemoto}
\affiliation{Philippe Meyer Institute for Theoretical Physics, Physics Department, \'Ecole Normale Sup\'erieure \& PSL Research University, 24 rue Lhomond, 75231 Paris Cedex 05, France}
\author{\'Etienne Fodor}
\affiliation{Department of Applied Mathematics and Theoretical Physics, University of Cambridge, Wilberforce Road, Cambridge CB3 0WA, United Kingdom}
\author{Michael E. Cates}
\affiliation{Department of Applied Mathematics and Theoretical Physics, University of Cambridge, Wilberforce Road, Cambridge CB3 0WA, United Kingdom}
\author{Robert L. Jack}
\affiliation{Department of Applied Mathematics and Theoretical Physics, University of Cambridge, Wilberforce Road, Cambridge CB3 0WA, United Kingdom}
\affiliation{Department of Chemistry, University of Cambridge, Lensfield Road, Cambridge CB2 1EW, United Kingdom}
\author{Julien Tailleur}
\affiliation{Laboratoire Mati\`ere et Syst\`emes Complexes, UMR 7057 CNRS/P7, Universit\'e Paris Diderot, 10 rue Alice Domon et L\'eonie Duquet, 75205 Paris cedex 13, France}

\date{\today}

\maketitle

\section{Introduction}

Active particles constitute an important class of
non-equilibrium systems, with examples ranging from bacteria to
synthetic colloidal
swimmers~\cite{Berg2008book,Cates2012RPP,Howse2007PRL,Theurkauff2012PRL,Palacci2013Science,Bricard2013Nature,BechingerRMP}. These
particles expend energy to propel themselves, driving
active matter out of equilibrium at microscopic scales and causing rich dynamical behaviors. Some of these are
\emph{universal}, whereby systems that differ microscopically
show similar emergent physics~\cite{Marchetti:2013:RMP}, such as motility
induced phase separation (MIPS)~\cite{Cates:2015:ARCMP,Tailleur:2008:PRL,FilyYaouenMarchettiCristina,RednerHaganBaskaran,Bialke:2013:EPL,Wysocki:2014:EPL,Stenhammar:2014:SM}, collective motion~\cite{Vicsek1995PRL,Gregoire2004PRL,Ballerini2008PNAS,Vicsek2012PhysRep,Solon2015PRLflock,Deseigne2010PRL,Bricard2013Nature,Deseigne2010PRL}, lane formation~\cite{Klymko2016,Junco2018} or motile defects~\cite{Narayan2007Science,Sanchez2012Nature,Giomi2013PRL,Thampi2014EPL,Decamp2015Nature,Mahault2018arXiv}.

Many recent advances in our understanding of non-equilibrium systems
are based on \emph{large-deviation theory}
(LDT)~\cite{Derrida2007JSM,Touchette_LD}. This extends the counting
procedures of equilibrium statistical mechanics from configuration
space to trajectory space, addressing collective phenomena such as
dynamical phase transitions. It has been used to characterize
dynamical
symmetries~\cite{Gallavotti1995PRL,Kurchan1998JPA,Lebowitz1999,Crooks1999PRE},
measure free energy
differences~\cite{Jarzynski1997PRL,Collin2005Nature}, and locate
atypical trajectories, such as activated
processes~\cite{Kohn2005JNS}. LDT has proven useful in fields ranging
from dynamical systems~\cite{Tailleur2007NatPhys} and
glasses~\cite{Garrahan2007PRL,hedges_dynamic_2009} to fluid
mechanics~\cite{Grafke2015JPA} and geophysical
flows~\cite{Bouchet2014FDR}. In contrast, the full range of insights
offered by LDT to active matter remains largely unexplored, despite a
handful of pioneering
studies~\cite{Thompson2011JSM,PhysRevLett.119.158002,whitelam2017phase,Chaudhuri2014PRE,Ganguly2013PRE,mandal2017entropy,shankar2018hidden}.

Here we use LDT to study active Brownian particles (ABPs) interacting
via repulsive central forces.  We focus on the large deviations of the
active work, defined as the particle-averaged inner product of
propulsive force and velocity. This measures how far the local
self-forcing of active particles translates into real motion. A recent
study~\cite{PhysRevLett.119.158002} used brute-force simulations to
sample the fluctuations of active work in a dilute system of active
dumbbells and found a low active work to correlate with the emergence
of ordered clusters in this system. Here we use an advanced numerical
method~\cite{Giardina_Kurchan_Peliti,Tailleur2007NatPhys,Hurtado2009PRL,Giardina2011JSP,Nemoto_Bouchet_Jack_Lecomte,Nemoto_Jack_Lecomte,ray2017exact,klymko2018rare,Brewer_Clark_Bradford_Jack}
to explore the full large-deviation regime in all relevant regions of
the phase diagram of our ABP model. We first show that finite systems
always admit a large deviation principle,
notwithstanding~\cite{PhysRevLett.119.158002}, but that
{they are flanked} by two dynamical
phase transitions. Indeed, minimizing the active work always leads to
complete dynamical arrest, whether or not the unbiased system exhibits
MIPS. Biasing instead towards high active work, we find a striking
result: trajectories now correspond to flocked states of aligned
collective motion, despite the microscopic absence of aligning
interactions.  We explain the origin of the dynamical phase
transitions separating these regimes using a combination of arguments
including macroscopic fluctuation
theory~\cite{Bertini2002JSP,Jack2015PRL}.

\section{Model}
We consider $N$ active Brownian particles interacting via purely repulsive pairwise forces in two spatial dimensions~\cite{FilyYaouenMarchettiCristina,RednerHaganBaskaran,Stenhammar:2014:SM,Wysocki:2014:EPL,Bialke:2013:EPL}.
The positions and orientations of the particles are $\bm{r}_i$ and $\theta_i$; they evolve as
\begin{align}
\label{eqr_i}
\dot {\bv r} _i &= \mu \bv F_{i,\rm ex} + \vp {\bv u}(\theta_i) + \sqrt{2 D} 
\bv \eta_i\; ; \quad
\dot {\theta_i}  =  \sqrt{2 D_{\rm r}} \xi_i\;,
\end{align}
where ${\bv \eta}_i, \xi_i$ are zero-mean unit-variance Gaussian white
noises, $\mu$ is a particle's mobility, $\vp$ its bare self-propulsion
speed and ${\bv u}(\theta_i) = (\cos\theta_i,\sin\theta_i)$ its
orientation vector, and $D, D_r$ are translational and
rotational diffusivities. Particles interact via a
repulsive WCA potential, detailed in Appendix~\ref{appendix:nondimensionalized}, of
range $\sigma$. {For consistency with
  \cite{RednerHaganBaskaran}, we set $D_r=3D/\sigma^2$ and the WCA
  strength parameter to be $D/\mu$. Then, we choose space and time units such that $\sigma=1$ and  
  $\sigma/\vp=1$ (see Appendix~\ref{appendix:nondimensionalized}). When the
persistence length $\ell_{\rm p}\equiv \vp/D_{\rm r}$ is much larger than the
particle size, $\ell_{\rm p}/\sigma\gtrsim 15$, the
system undergoes MIPS: at high volume fractions, a vapor of motile particles coexists
with dense macroscopic
clusters~\cite{FilyYaouenMarchettiCristina,RednerHaganBaskaran,Wysocki:2014:EPL,Stenhammar:2014:SM,Bialke:2013:EPL}. For
smaller $\ell_{\rm p}$, the system remains uniform and the main effect of 
activity is to enhance the effective translational diffusivity.

For interacting particles, a natural measure of how efficiently active forces create motion is given by the propulsive speed $v_i\equiv \dot{\bv r_i} \cdot {\bv u(\theta_i)}$ which projects a particle's velocity along its orientation. This relates directly to the active work~\cite{PhysRevLett.119.158002},  the total work done by the active forces on the particles, which obeys (in the Stratonovich convention)
\begin{equation}
\Wact(t) \equiv  \frac{\vp}{\mu} \int_{0}^{t}  \sum_{i=1}^{N}\dot {\bv r}_{i}(\tau) \cdot {\bv u}\big(\theta_i(\tau)\big) \, \mathrm{d}\tau=\sum_{i=1}^{N} \int_{0}^{t} \frac{\vp v_i(\tau)}{\mu} \, \mathrm{d}\tau.
\label{equ:Wact}
\end{equation}
For conservative interactions, $\Wact$ relates to the
dissipation in the thermostat $\int dt \dot{\bv r}_i\cdot(\dot{\bv
  r}_i-\sqrt{2D}{\bv\eta}_i)$~\cite{Sekimoto1997, Toyabe2010,
  Seifert2012, Kanazawa2014, Ahmed2016}, and thus to the entropy
production in the full $\{r_i,\theta_i\}$ configuration
space~\cite{PhysRevLett.119.158002}. (This is generally distinct from
that measured in position space
$\{r_i\}$~\cite{PhysRevLett.117.038103}. {See also~\cite{Puglisi2017,marconi2017heat} for a comparative study of different entropy productions.})
It is convenient to consider a normalized rate of
active work per particle, $w \equiv \Wact \mu/(v_p^2 N t)$.
The dilute limit of vanishing packing fraction $\phi\to0$  then leads to $\langle
w\rangle=1$ which serves as a useful reference point. 

For fixed $N$ and large $t$}, the distribution of $w$ has a
large-deviation form
\begin{equation}
p(w) \sim \exp [- t I(w)]\;,
\label{equ:ldp-w}
\end{equation}
where $I(w)$ is a rate function~\cite{Touchette_LD}. 
The corresponding
 cumulant generating function (CGF)
\begin{equation}\label{eq:Gofs}
G(s) = \lim_{t \rightarrow \infty} \frac{1}{t} \log  \left \langle e^{-st N w} \right \rangle\; 
\end{equation}
is related to $I(w)$ by Legendre transformation.  As shown in Appendix~\ref{appendix:fluctuationrelation}, the functions $I(w)$ and $G(s)$ are convex, and $G(s)$
obeys a fluctuation relation $G(s) = G(a-s)$, with $a=3\ell_p/\sigma$.
The CGF is analogous to a free energy in equilibrium statistical mechanics~\cite{lecomte2007thermodynamic,
Touchette_LD,garrahan_first-order_2009,chetrite2013nonequilibrium}.  
Within this analogy, \emph{trajectories} of our two-dimensional system (evolving in time) correspond to \emph{configurations} of an anisotropic three-dimensional system.  Suppose that one spatial dimension (the ``length'') of this anisotropic system 
becomes infinite, while the others remain fixed -- this is analogous to considering trajectories with $t\to\infty$ and fixed $N$.
Phase transitions are not possible in such one-dimensional geometries, which is another way to see that $G(s)$ must be convex (and analytic).

\begin{figure}
\begin{center}
\hspace{-0.9cm}
\includegraphics[width=70mm]{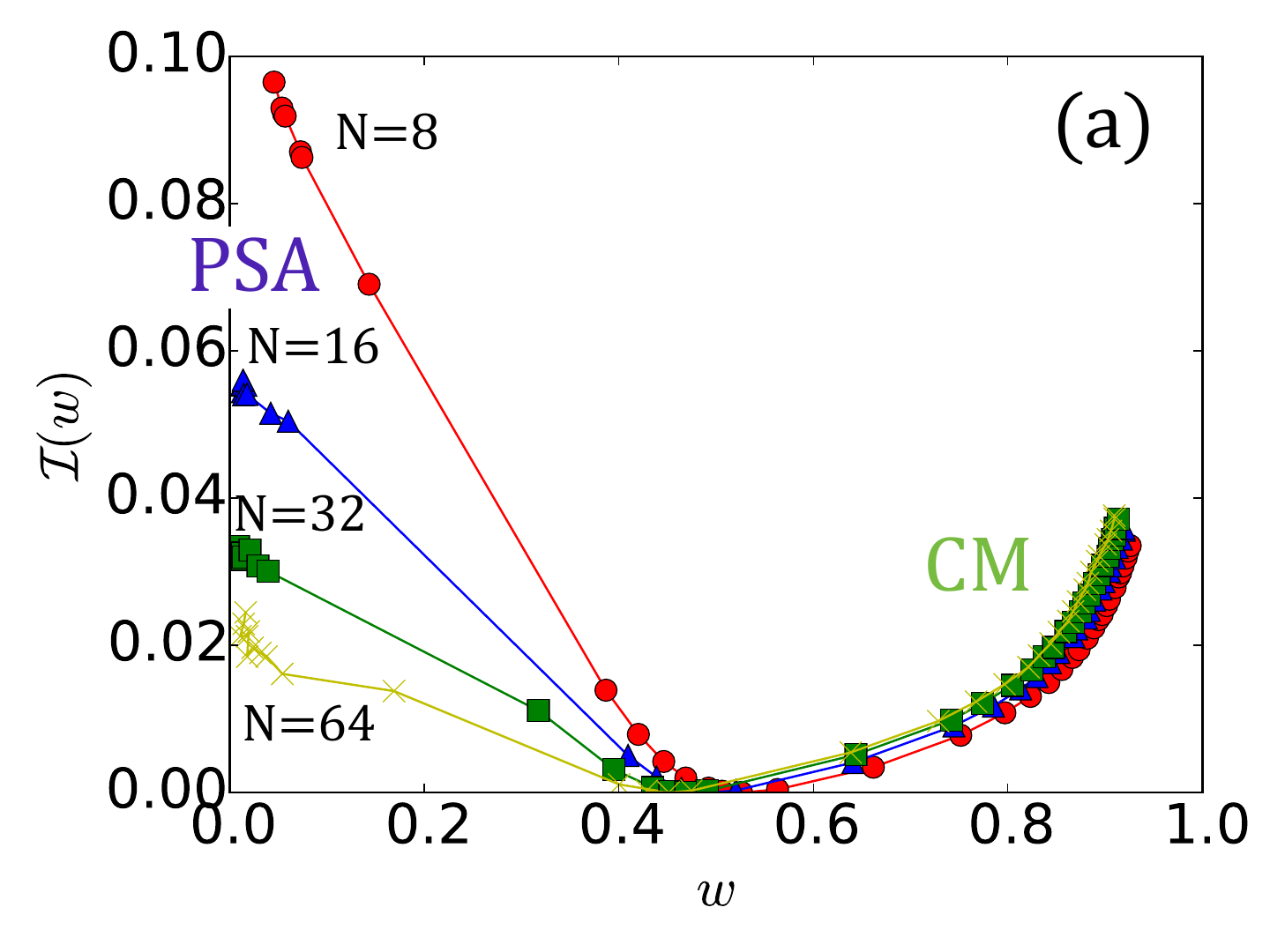}   \\ 
\hspace{-0.9cm}
\includegraphics[width=70mm]{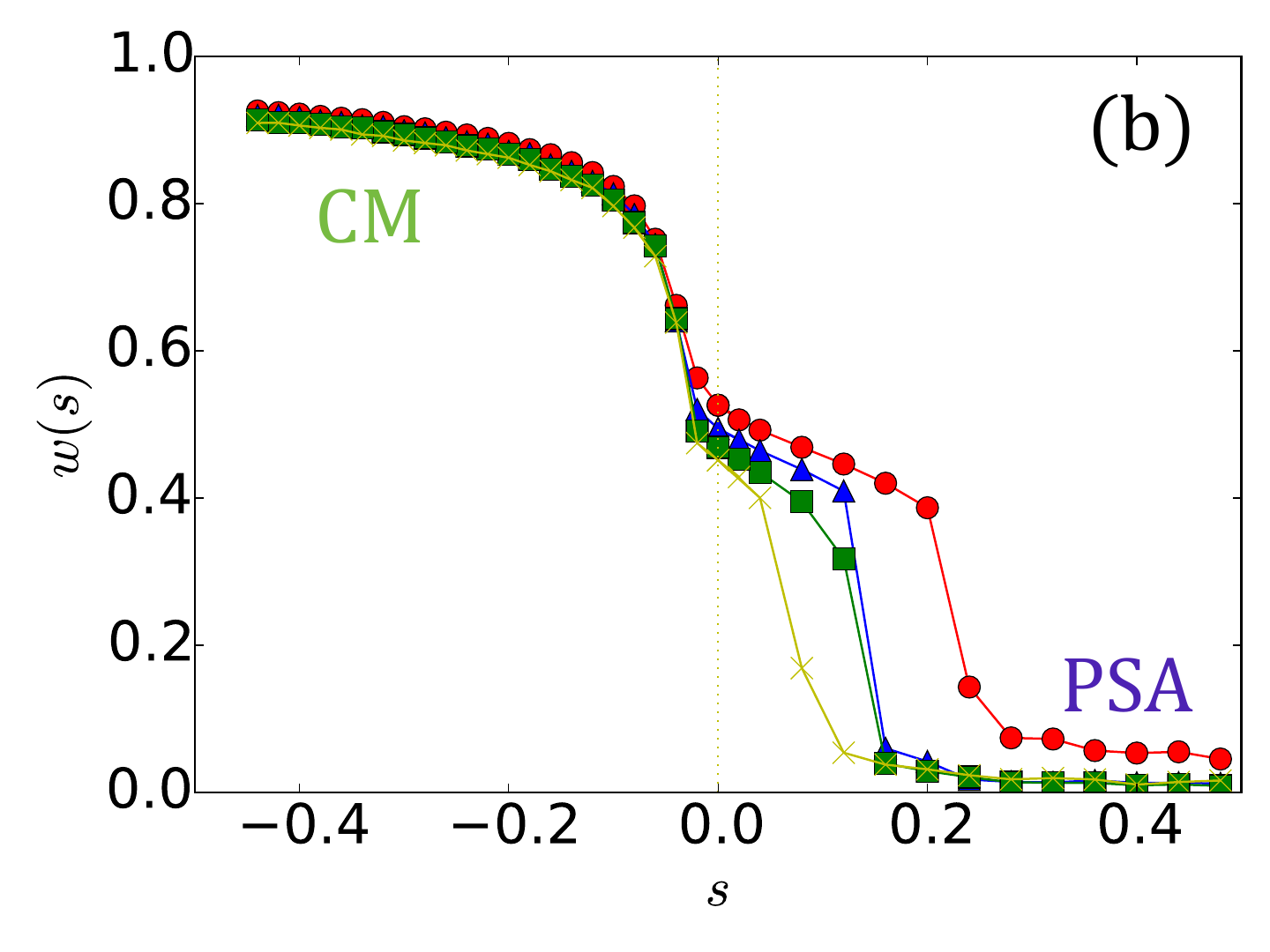} \\
\includegraphics[width=42mm]{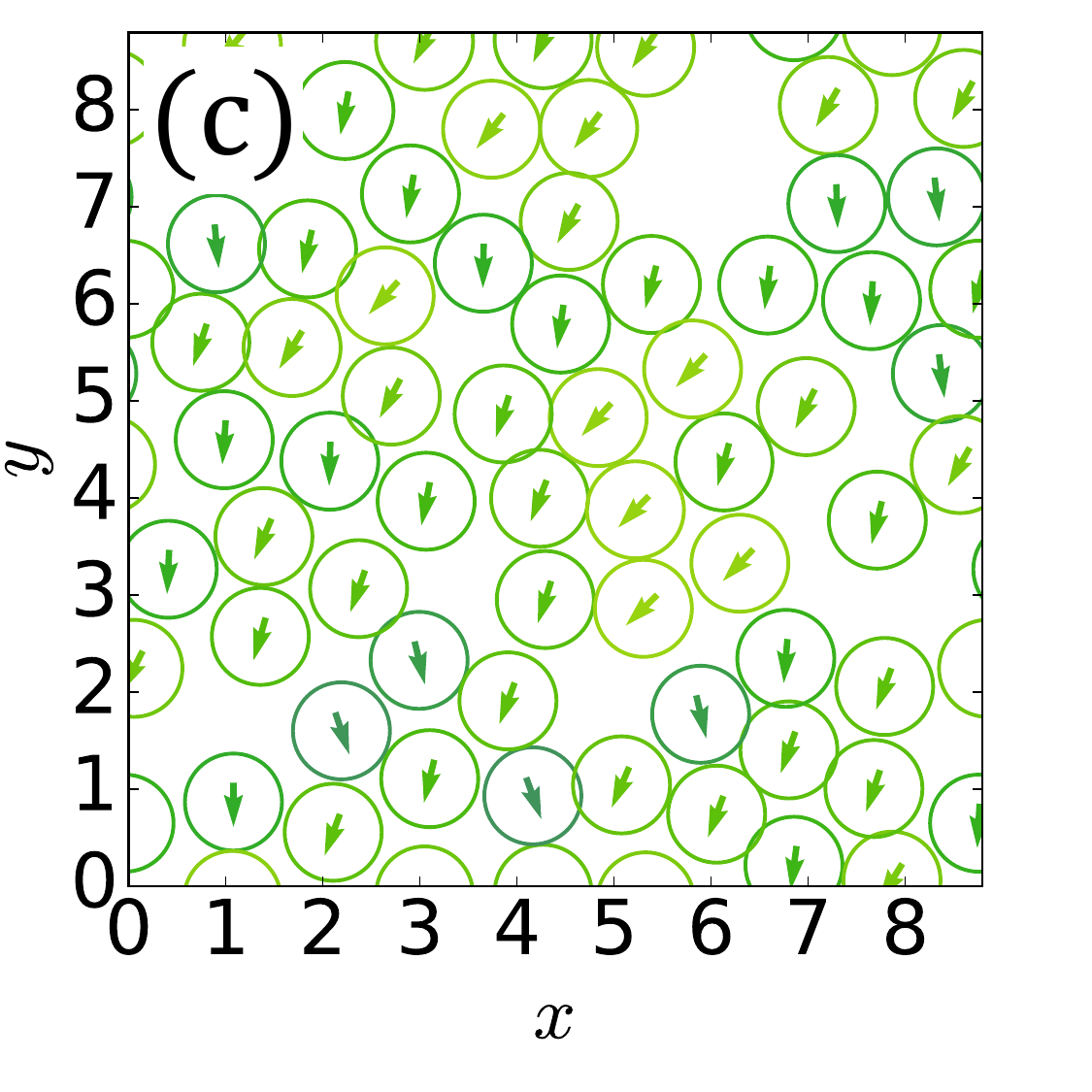}
\includegraphics[width=42mm]{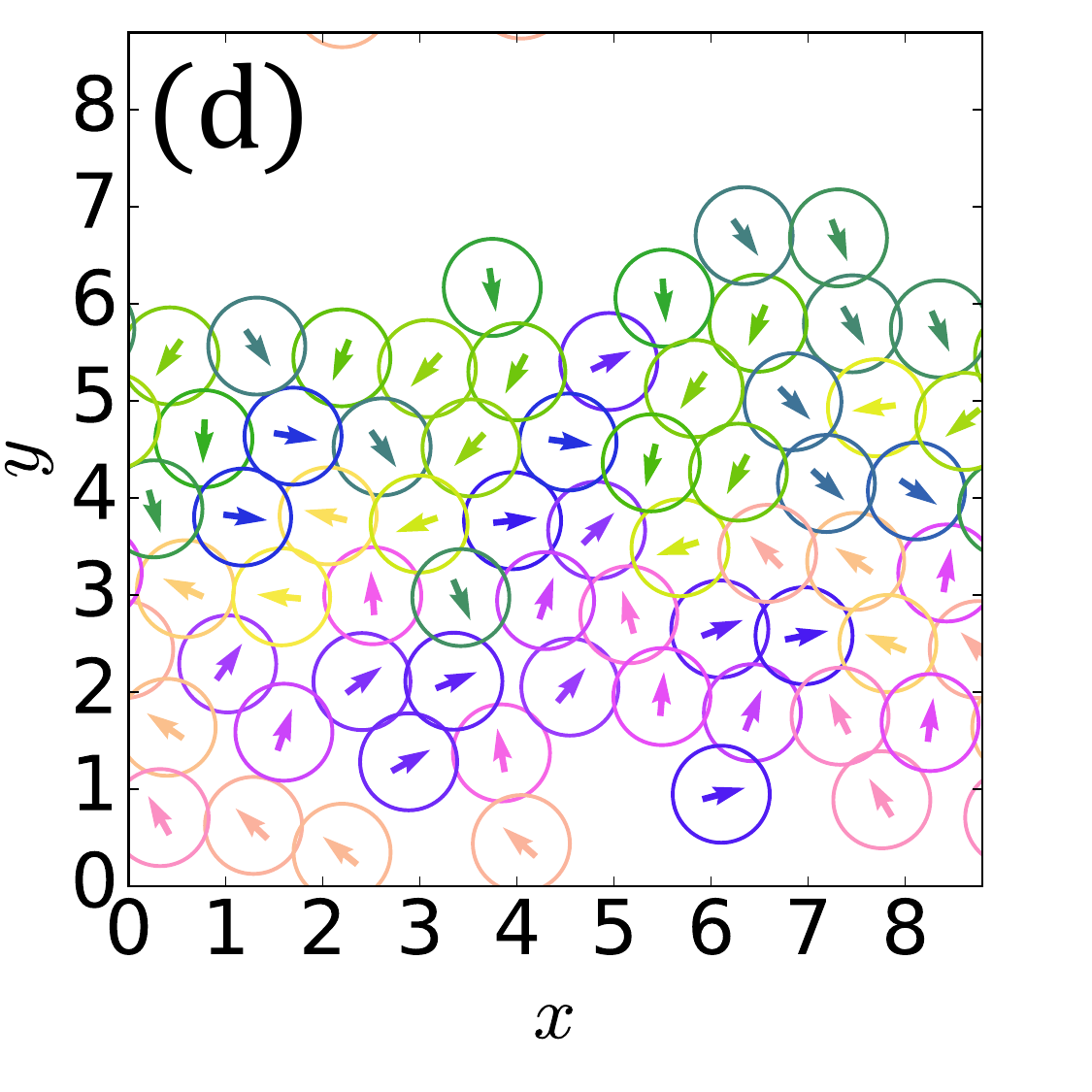}
\caption{ {\bf (a)} The rescaled rate function ${\cal I}(w)$, for system parameters where MIPS occurs ($\phi=0.65$, $\ell_p=40\sigma$).  The $N$-dependence shows strong finite size effects for $w<\langle w\rangle$.  {The labels PSA and CM indicate respectively regimes of phase-separated arrest and collective motion.}
{\bf (b)} The function $w(s)$ exhibits two sharp crossovers, which we attribute to two dynamical phase transitions.  (These appear in (a) as near-linear segments of the rate function.)  
{\bf (c,d)} Snapshot configurations for $N = 64$ in the biased ensemble, corresponding to the arrested phase
  ($s=0.8$, {(d)}) and the collective motion phase ($s=-3.2$,
  ${(c)}$). Particles are colored according to their orientations. For corresponding movies see~\cite{Suppl}. }\label{fig:RateFunction}
\end{center}
\end{figure}

Now consider the 
limit $N\to\infty$ (taken at fixed $\phi$, after $t\to\infty$).  In this case dynamical phase transitions are possible -- the analogous thermodynamic system is becoming infinite along more than one spatial dimension~\cite{garrahan_first-order_2009,jack2010large,Nemoto_Jack_Lecomte}.
The dynamical analogues of the (bulk) thermodynamic free-energy and entropy are 
\begin{equation}
  {\cal I}(w)=I(w)/N\qquad \text{and} \qquad{\cal G}(s)=G(s)/N\; .
\end{equation}
As in statistical mechanics, singularities in the large-$N$ limits of these functions are interpreted as phase 
transitions~\cite{bodineau2005,bertini2005,garrahan_first-order_2009,jack2010large,Nemoto_Jack_Lecomte,vaik2014}.

To observe and measure large deviations of the active work, we use a
cloning
algorithm~\cite{Tailleur2007NatPhys,Giardina2011JSP,Nemoto_Bouchet_Jack_Lecomte},
also known as Population Monte Carlo~\cite{delmoral}, whose
optimised implementation using modified dynamics~\cite{Nemoto_Bouchet_Jack_Lecomte} is detailed
in Appendix~\ref{appendix:enhanced_convergence}. (See~\cite{Giardina_Kurchan_Peliti,Brewer_Clark_Bradford_Jack}
for a lattice version of this algorithm, and~\cite{whitelam2017phase}
for a recent application to active systems.) In essence, the method
relies on evolving a large population of copies of the system to
generate ``biased ensembles'' of trajectories that sample the average
in~\eqref{eq:Gofs} {with a cost that scales linearly in $t$, allowing direct access to the large-$t$ limit}. For positive and negative $s$, {the biased ensembles} 
are dominated
by trajectories with atypically small and large $w$ respectively.

\section{results}

We first consider a system whose parameters lie (as $N\to\infty$)
within the MIPS region, $\ell_p=40 \sigma$ and $\phi=0.65$
(See~\cite{RednerHaganBaskaran} for the full phase diagram of the
system). We compute ${\cal G}(s)$, and $w(s)\equiv -{\cal G}'(s)$,
which is the mean value of the active work in the presence of the
bias, and its inverse $s(w)$.  We also determine the rescaled rate
function as ${\cal I}(w) = - s(w) w - {\cal G}(s(w))$.  Our numerical
results (Fig.~\ref{fig:RateFunction}) show three regimes separated by
dynamical phase transitions that we discuss below: a MIPS-like
coexistence between vapor and dense phases near $s=0$; a
phase-separated arrest (PSA) at large positive $s$; and a collectively
moving (CM) state at large negative $s$.

\begin{figure}
\begin{center}
\hspace{-0.8cm}
\includegraphics[width=65mm]{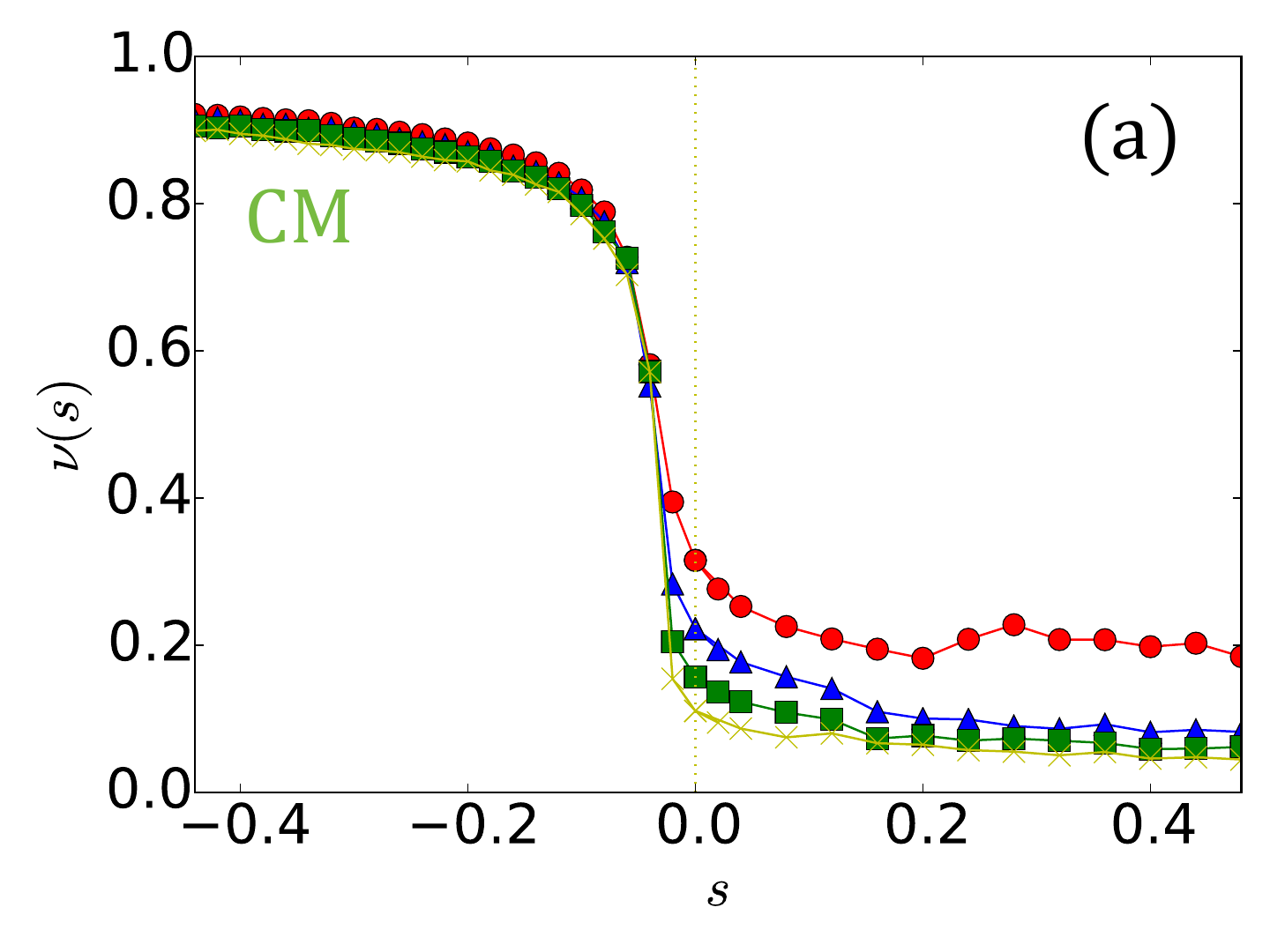} \\
\hspace{-1cm}
\includegraphics[width=70mm]{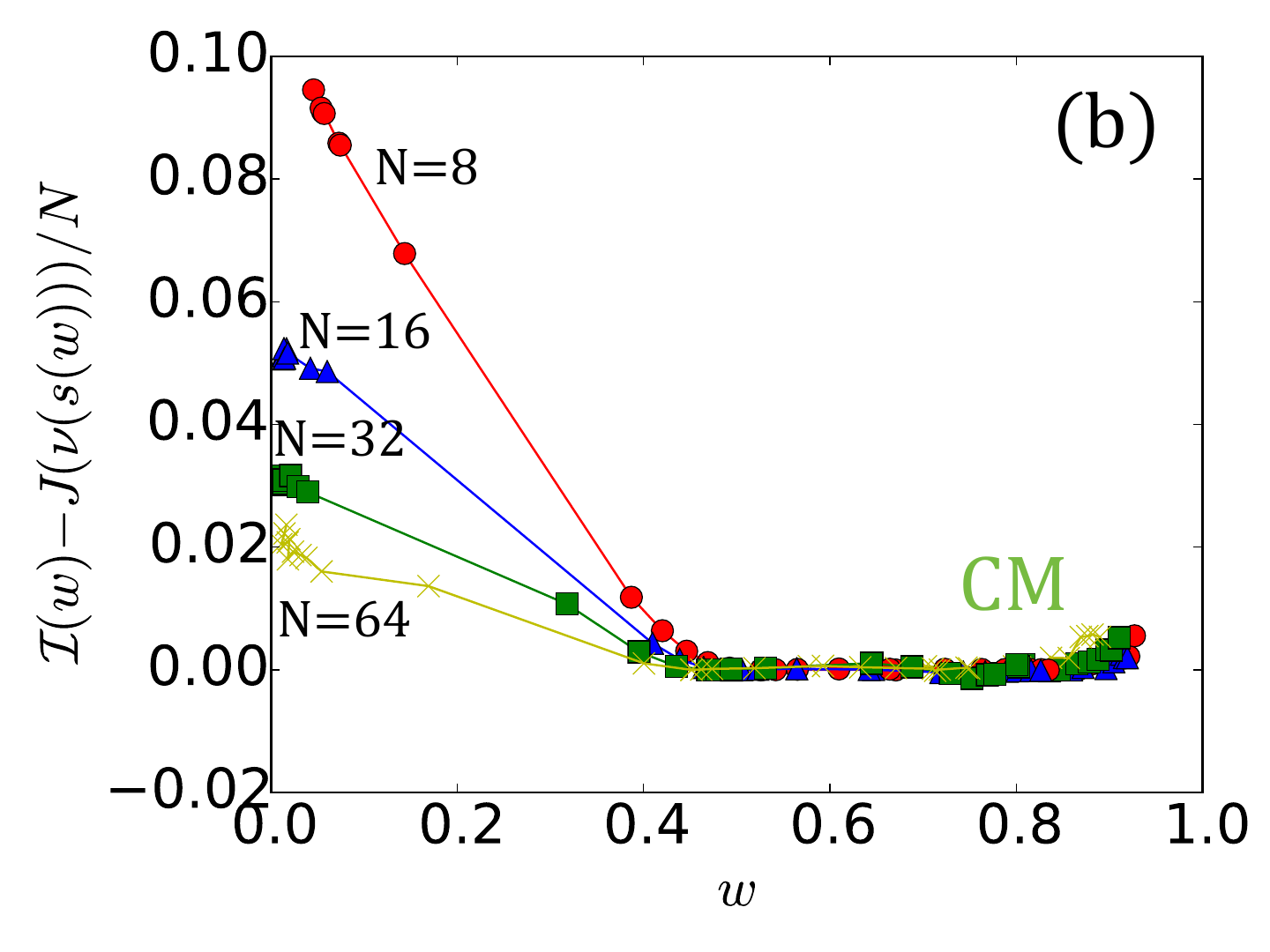}
\caption{\label{fig:orientation} {\bf (a)} The orientation vector
  $\nu(s)$ in the biased ensemble.  {\bf (b)} The difference between
  the two quantities appearing in (\ref{equ:IJ}), showing saturation
  of this bound at $w>\langle w\rangle_0$.
  }
\end{center}
\end{figure}

\subsection{Large active work: Collective motion} 
For large negative $s$, the
biased ensemble probes atypically
large values of the active work. Despite the absence of
aligning interactions, the biased ensemble is dominated by
trajectories where particles' orientations are aligned with each
other, and they move collectively as a flock.  A global order parameter for this
transition is the orientation
\begin{equation}
\hat\nu = \bigg| \frac{1}{N} \sum_{i=1}^{N} {\bv u}(\theta_i) \bigg| , \qquad \nu(s) = \langle \hat\nu \rangle_s\,,
\label{eq:defnu}
\end{equation}
where angle brackets are averages within the biased
ensemble. Fig.~\ref{fig:orientation} shows
the emergence of global orientation for $s<0$. In contrast, in
the unbiased case ($s=0$) there is no emergent alignment:  $\nu(s=0)\to 0$ at large $N$.

To understand the emergence of order, note that 
{particle alignment naturally promotes}
active work: the CM state has far fewer collisions than if motion is incoherent,
so that active forces translate more efficiently into particle motion. To confirm this
interpretation, we compute
the rate function $J(\bar\nu)$ of the time-integrated orientation
$\bar\nu = t^{-1} \int_0^t \hat\nu(\tau) \mathrm{d}\tau$ whose
probability distribution scales as $p(\bar\nu) \sim {\rm
  e}^{-tJ(\bar\nu)}$, analogous to (\ref{equ:ldp-w}). 
Since the
orientational dynamics of ABPs is independent of their positions, the rate function $J$ can be computed semi-analytically as shown in Appendices~\ref{appendix:derivation} and \ref{appendix:cgf}.
It measures the probability of rare events where rotational symmetry is spontaneously violated.
Now, define
$I_2$ as the joint rate function for $\bar\nu$ and $w$, and let
$\nu_*(w)=\nu(s(w))$ be the average global orientation for a biased
system with active work $w$.  Then $I(w) = \inf_\nu I_2(w,\nu) =
I_2(w,\nu_*(w))$ where the first equality is the contraction principle
for large deviations~\cite{Touchette_LD} and the second follows
because the infimum is achieved by $\nu_*$.  Similarly $J(\nu_*(w)) =
\inf_{w'} I_2(w',\nu_*(w)) \leq I_2(w,\nu_*(w))$ and hence
 \begin{equation}
 {\cal I}(w) \geq \frac{J(\nu_*(w))}{N} .
 \label{equ:IJ}
 \end{equation}

Fig.~\ref{fig:orientation}(b) shows that the inequality (\ref{equ:IJ})
is almost saturated when $w>\langle w\rangle_0$.  Physically, this
indicates that the probability cost for creating a large fluctuation
of the active work is dominated, for $s<0$, by the cost to create an
improbable global orientation, with an associated spontaneous symmetry
breaking and an accompanying singularity in ${\cal I}(w)$.  It appears
that ``the best strategy'' for a set of active particles to move fast
is for them to collectively align.  Here ``best'' means least improbable
within the microscopic stochastic dynamics specified by
\eqref{eqr_i}. {Note that the emergence of macroscopic arrested
  clusters due to MIPS can be suppressed by local torques that limit
  the head-on collisions of
  particles~\cite{pohl2014dynamic,matas2014hydrodynamic,zottl2014hydrodynamics}. It
  is thus quite remarkable that the most likely way to generate an
  efficient motion of each particle, and hence a large active work, is
  through the emergence of a collectively moving state, and not
  through such local rearrangements 
{(which do not lead to a CM state)}.}

\subsection{Small active work: Dynamic arrest} 
For positive $s$, the
biased ensemble selects trajectories with atypically low active work,
so that propulsive effort leads to little motion.  On increasing the
bias, we find the system sharply transitions into a dynamically
arrested state. (See Fig~\ref{fig:RateFunction}c and movies
in~\cite{Suppl}). A signature of this transition is the discontinuity
in $w(s)$ reported in Fig~\ref{fig:RateFunction}(b), which signifies a
first-order dynamical phase transition: the linear segment in ${\cal
  I}(w)$, for $w<0.5$, is analogous to a Maxwell construction and the
discontinuity in $w(s)$ to a jump in the order parameter. These
features should become strict singularities only as $N\to\infty$ but
are clearly visible for $N = 32,64$. (Note however that $G(s)$ and $I(w)$ are
well-defined for any finite $N$,
notwithstanding~\cite{PhysRevLett.119.158002}.)  As $N$ increases, the
critical value of $s$ moves towards zero (see Appendix~\ref{appendix:Finite-size} for this
finite-size scaling analysis), suggesting that bulk MIPS states live
on the verge of a first order transition to complete arrest.

This situation is reminiscent of dynamical phase transitions arising
for activity-biased kinetically constrained models (KCMs) of glassy
systems~\cite{Garrahan2007PRL,garrahan_first-order_2009}. Indeed, our
findings for $s>0$ can be qualitatively understood by generalizing
arguments developed for KCMs.  {Specifically, we exploit a variational
  principle that allows the rate function $I(w)$ to be computed by
  considering what auxiliary `control' forces need to be added to a
  system to realise the rare trajectories of
  interest~\cite{Nemoto2011,TouchetteJSM2015,JackSollichEPJ2015}.}
Stabilizing a large, dense cluster in a system undergoing MIPS only
requires applying forces on its boundaries, hence involving a
sub-extensive number of particles. This argument, detailed
in Appendix~\ref{Appendix:optimal_variational}, immediately leads to $\lim_{N\to\infty}{\cal I}(w)=0$
and a dynamical phase transition at $s=0$.  Since the argument is
  variational in nature, it can be exploited in numerical simulations:
  we have used it to obtain bounds on $I(w)$ for $w<\langle w\rangle$
  in large systems as shown in Appendix~\ref{Appendix:optimal_variational}, which are consistent with the presence of a phase
  transition and complement the accurate results for $I(w)$ in small
  systems that we show in Fig.~\ref{fig:RateFunction}.

\subsection{Large deviations for homogeneous steady states} 
\begin{figure*}
\begin{center}
\includegraphics[width=70mm]{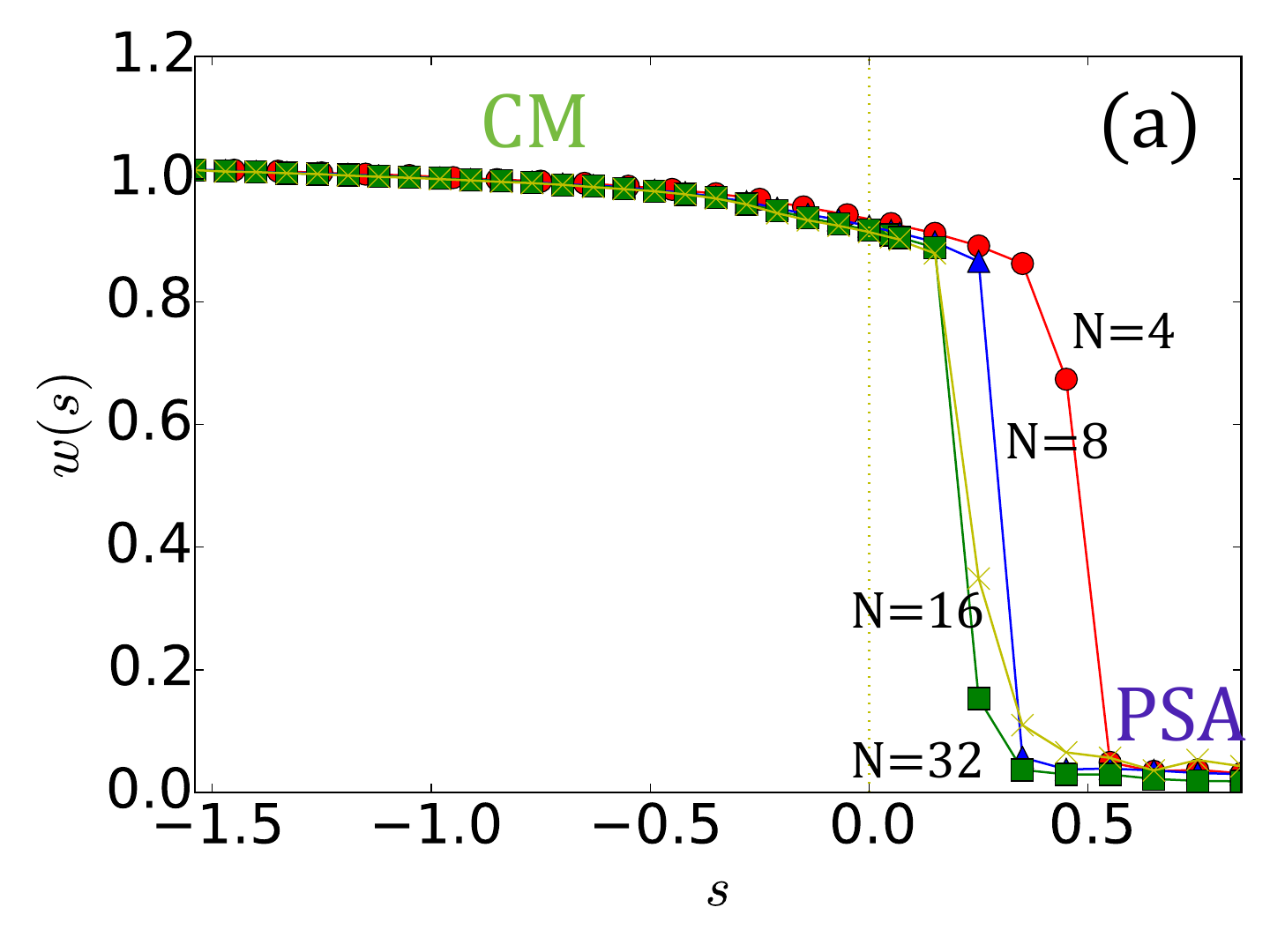}
\includegraphics[width=70mm]{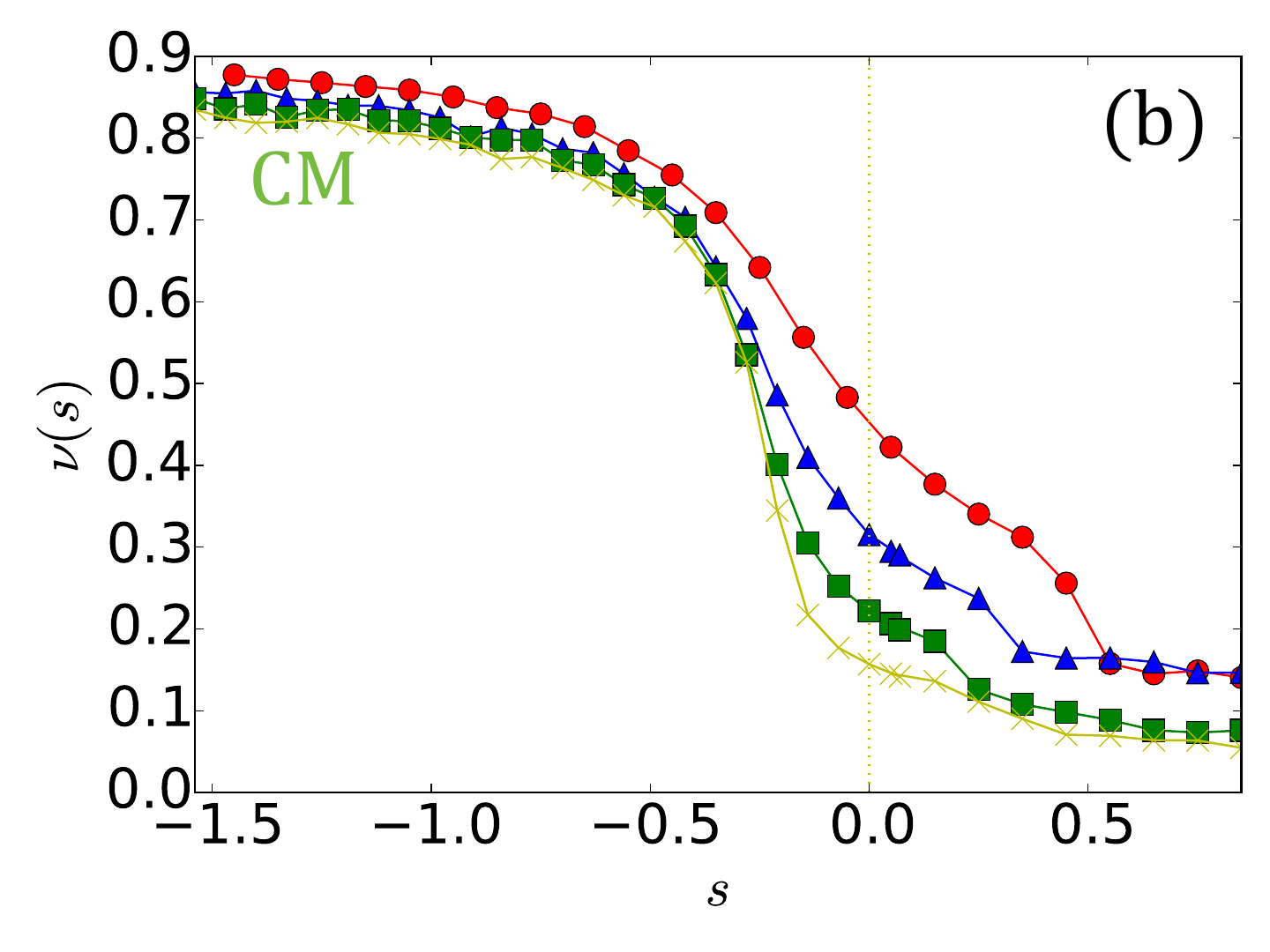}
\includegraphics[width=70mm]{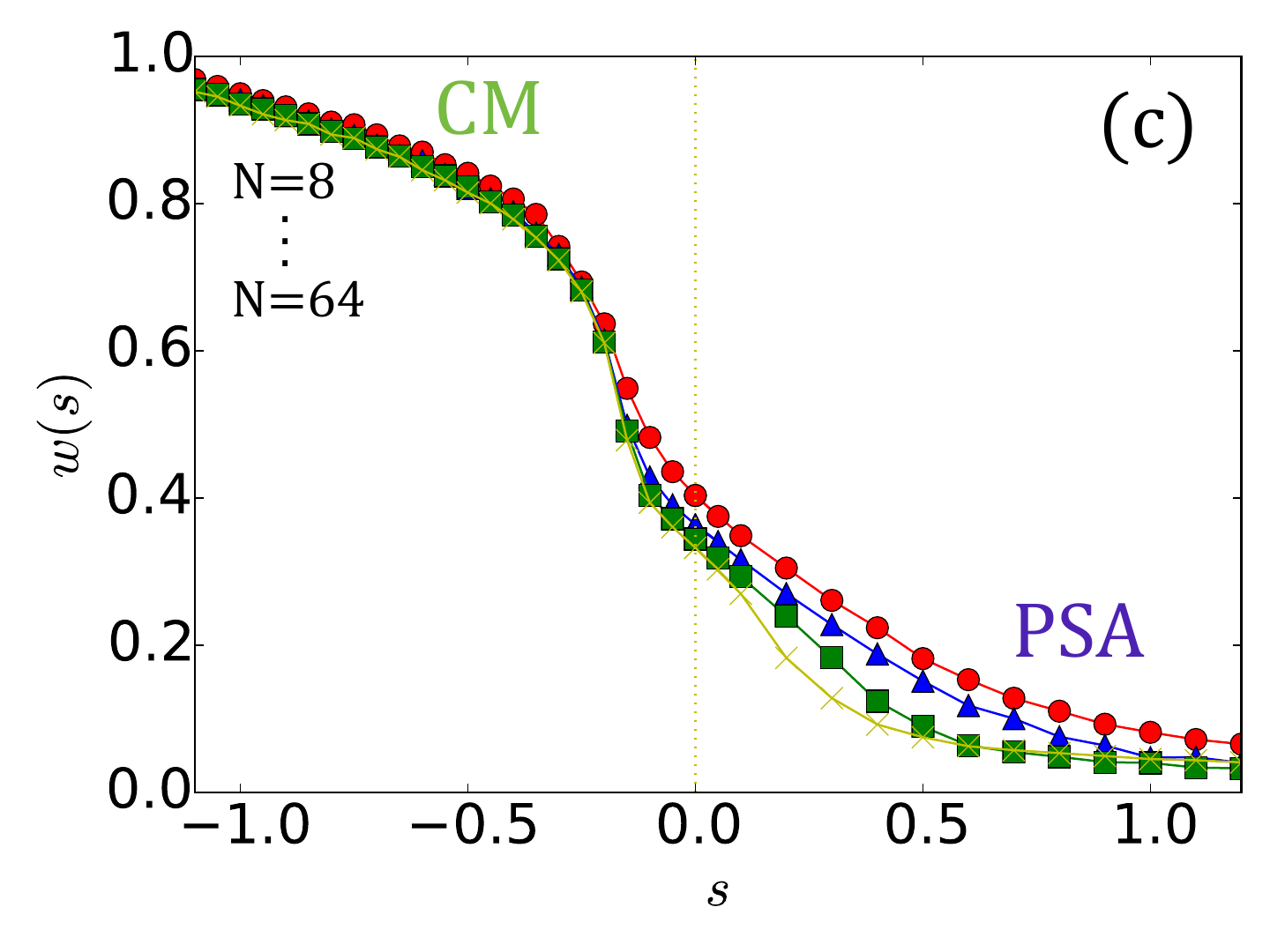}
\includegraphics[width=70mm]{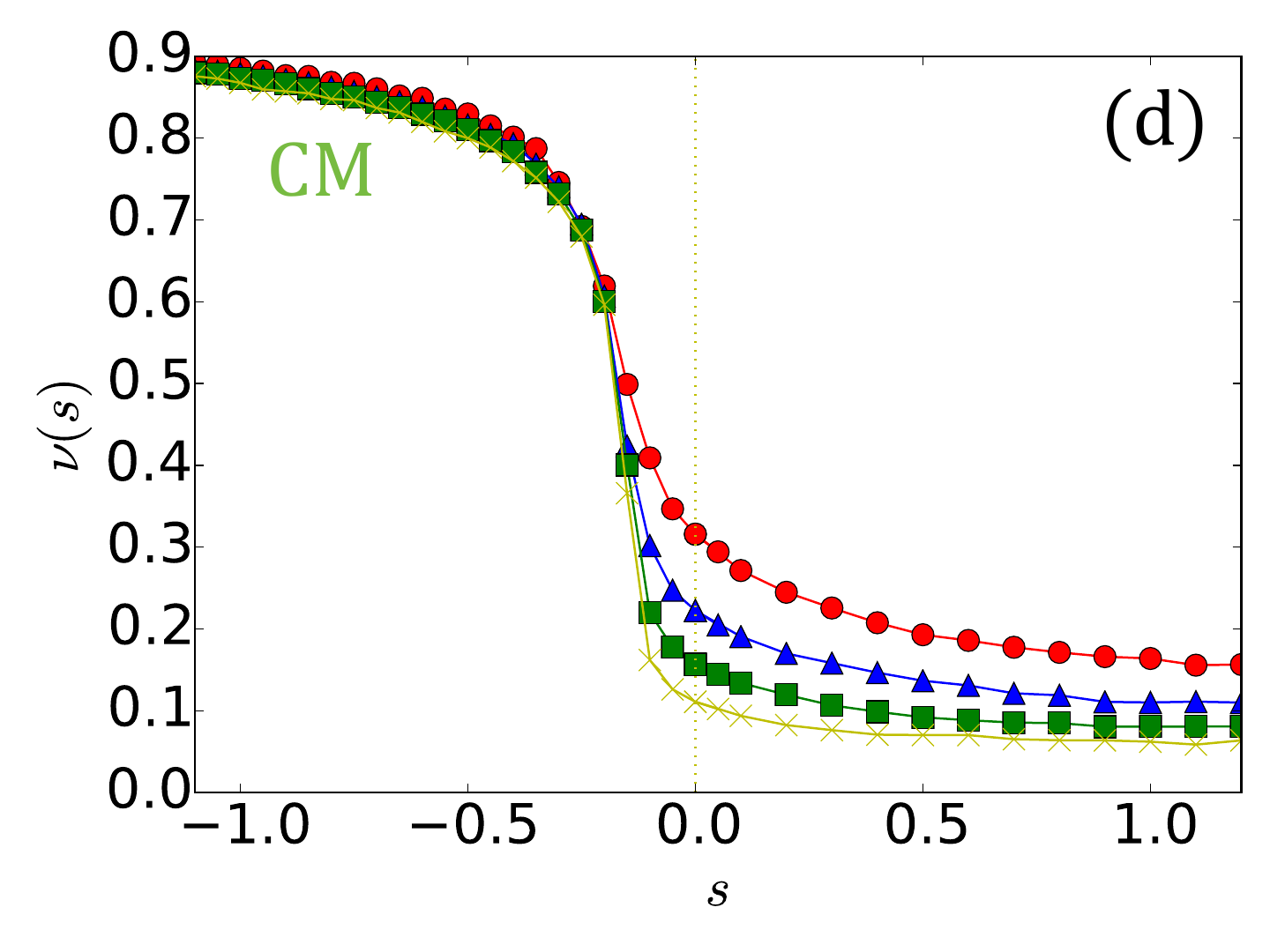}
\caption{\label{fig:Ptail_2} 
The active work $w(s)$ and the orientation $\nu(s)$ for two state points where the unbiased dynamic gives homogeneous steady states.  In (a,b) the state point is $(\phi,\ell_p) = (0.1,40\sigma)$; in (c,d) it is $(\phi,\ell_p) = (0.65,6.7\sigma)$. The crossovers shown in these figures separate the homogeneous fluid ($s=0$) from a CM phase ($s<0$) and a PSA state ($s>0$). These phases are qualitatively similar to those shown in Fig \ref{fig:RateFunction}(c,d). See movies in \cite{Suppl}.}
\end{center}
\end{figure*}

So far we considered systems with parameters for which the unbiased, large $N$ dynamics shows steady-state MIPS. We now consider large deviations from a steady state that is homogeneous. We focus on two state points: $(\phi,\ell_p) = (0.1,40\sigma)$, corresponding to a reduced density but large $\ell_p$, and $(\phi,\ell_p) = (0.65,6.7\sigma)$, corresponding to high density but smaller $\ell_p$. Fig.~\ref{fig:Ptail_2} and movies in~\cite{Suppl} show  the asymptotic phases observed for large positive and negative $s$ to be similar in both cases: for $s<0$, they again exhibit collective motion while for $s>0$ the system undergoes phase-separated arrest.   
{Compared to Fig.~\ref{fig:RateFunction}, the crossover to the PSA state in Fig.~\ref{fig:Ptail_2}c is smoother; at this smaller value of $\ell_p$, the density fluctuations of the ABPs are smaller (there is no MIPS) and the instability to phase separation is weaker.}

This bias-induced phase separation can be explained by a hydrodynamic argument~\cite{Jack2015PRL}. For a macroscopically homogeneous fluid, long-wavelength density fluctuations should obey an equation of the form~\cite{Bertini2015}
\begin{equation}
\dot\rho = \nabla \cdot \left( D_{\rm eff}(\rho) \nabla\rho + \sqrt{2\sigma(\rho)} \xi \right)
\label{equ:mft}
\end{equation}
where $\rho$ is the local density, $D_{\rm eff}$ is a
(density-dependent) diffusivity, $\sigma(\rho)$ is a noise strength,
and $\xi$ is a Gaussian white noise.  (Higher-order gradients, while
relevant to
MIPS~\cite{wittkowski2014Ncom,nardini2017PRX,solon2018PRE,tjhung2018arxiv},
are negligible for the long-wavelength fluctuations of interest here.)
It is then natural to approximate $W[\rho]\simeq \frac{v_{\rm p}}\mu
\int dt dx \rho v(\rho)$, where $v(\rho)$ is the average of the
effective active speed $v_i$ in a homogeneous system of density
$\rho$. This $v(\rho)$ is known to decrease linearly with density in
pairwise-force active particles~\cite{FilyYaouenMarchettiCristina} so
that the active work density $\kappa(\rho)\equiv \frac{v_{\rm p}}\mu
\rho v(\rho)$ is a concave function. A density fluctuation $\delta
\rho$ then leads to a fluctuation of the active work $\delta W = \int
\frac{1}{2} \kappa''(\rho) (\delta\rho)^2 \mathrm{d}x$ with
$\kappa''(\rho)<0$.

Large deviations of such observables in the setting of~\eqref{equ:mft}
are known to lead to phase separation in the large system limit
$L\to\infty$ whenever $s>0$ and $\kappa''(\rho)<0$~\cite{Jack2015PRL}: a long-wavelength
linear instability arises for $s>\lambda_c/L^2$ with $\lambda_c =
(2\pi D_{\rm eff})^2/|\sigma
\kappa''|$~\cite{Appert2008,Lecomte2012,Jack2015PRL}. 
This bias-induced instability arises in passive systems~\cite{Lecomte2012,Jack2015PRL}, and we argue that it applies to homogeneous, isotropic active fluids also, since the form of (\ref{equ:mft}) is the same.  Alongside it, any conventional phase separation, including MIPS, creates an instability even in the unbiased case, $s = 0$. This sets in as $D_{\rm eff}(\rho)\to0$.  In that limit,  $\lambda_c\to0$ so that the bias-induced and motility-induced instabilities merge; physically, the bias reinforces the natural tendency to phase separate.  
{(The convergence
with $N$ is slowest in the small persistence length region,
Fig.~\ref{fig:Ptail_2}c, which is furthest from the MIPS regime.)}
In contrast, the collective motion regime observed for $s<0$ has no passive counterpart and cannot be captured by (\ref{equ:mft}), which assumes that the orientations are only weakly affected by the bias, and can therefore be integrated out.

\section{Conclusion} 
We have shown, using a combination of numerical simulations and theoretical arguments, that active systems interacting via pairwise forces undergo several different dynamical phase transitions. 
Choosing a bias field to select trajectories of low active work, we found these trajectories to involve a coexistence of a dense jammed, arrested domain with a dilute vapor. This is the most likely way in which an active system that is normally a uniform bulk fluid can stop moving.  Biasing in the other direction to find trajectories of high active work, we found collective motion with aligned propulsion directions despite the absence of aligning interactions microscopically. 

{We end by speculating about a link between large deviations and evolutionary biology, motivated by two observations.  
First, the cloning algorithm involves the \emph{evolution} of a \emph{population} of systems: the method balances their natural dynamics (which favour the
unbiased steady state) and a \emph{selection pressure}, which favours systems
with atypical values of some \emph{fitness} function (here,
active work)~\cite{brotto2016population,brotto2017model}.  
Second, we have shown that alignment among ABPs tends to suppress collisions, leading to efficient motion.  We have argued that alignment is an effective strategy for 
promoting particle motion, with a minimal cost (in probability).
We suggest that this cost-minimisation strategy might also be viewed as a possible \emph{evolutionary strategy} for maximising active work in biological systems.  
We do not expect a general correspondence between evolutionary strategies and cost minimisation, particularly since cost-minimisation strategies may be complicated, perhaps requiring concerted 
 motion across large length scales~\cite{jack2010large,Nemoto2011,chetrite2013nonequilibrium}.  However, one may imagine that some robust characteristics
(such as global alignment) might appear generically in \emph{both} cost-minimisation strategies \emph{and} evolutionary strategies.
}

\begin{acknowledgements}
This work was granted access to the HPC resources of CINES/TGCC under the allocation 2018-A0042A10457 made by GENCI and of MesoPSL financed by the Region Ile de France and the project Equip@Meso (reference ANR-10-EQPX-29-01) of the program Investissements d'Avenir supervised by the Agence Nationale pour la Recherche. \'EF benefits from an Oppenheimer Research Fellowship from the University of Cambridge, and a Junior Research Fellowship from St Catherine's College. JT acknowledges support from the ANR grant Bactterns. Work funded in part by the European Research Council under the Horizon 2020 Programme, ERC grant agreement number 740269. MEC is funded by the Royal Society.
\end{acknowledgements}

\appendix

\section{Non-dimensionalized time evolution equations and Peclet number}
\label{appendix:nondimensionalized}

We use the particle radius $\sigma$ and propulsion speed $v_{\rm p}$
to define dimensionless position and time as $\bv r_i/\sigma$ and
$v_{\rm p} t/\sigma$. We also define a dimensionless mobility
$\alpha=D/(D_{\rm r} \sigma^2)$. The dynamics
\eqref{eqr_i} then become
\begin{eqnarray}
\dot {\bv r} _i &=& \alpha \frac{ \sigma}{\ell_{\rm p} }\tilde { \bv F}_{i,\rm ex}  +   {\bv u}_i(\theta_i) + \sqrt{\alpha\frac{2\sigma }{\ell_{\rm p} }} \bv \eta _i,
\label{eqr_i_2}\\
\dot {\theta_i} &=&  \sqrt{\frac{2\sigma}{\ell_{\rm p} }} \xi _i.
\label{eqtheta_i_2}
\end{eqnarray}
The interaction force $\tilde { \bv F}_{i,\rm ex}$ stems from the
(dimensionless) WCA potential
\begin{equation}
  V_{\rm ex} = \big[ 4 \big((1/r)^{12} - (1/r)^6 \big)
+ 1 \Big]\Theta(2^{1/6}-r)
\end{equation}
where $\Theta$ is a Heaviside step
function. (Following~\cite{RednerHaganBaskaran}, we chose the typical
strength of the WCA potential to be $D/\mu$ in the original units.)
The Peclet number $\rm Pe$ used in~\cite{RednerHaganBaskaran} is given
as
\begin{equation}
{\rm Pe}=\frac{\ell_p}{\sigma \alpha}
\end{equation}
with $\alpha = 1/3$. The normalized active work rate is
$w(t) = (1/(Nt))\sum_{i=1}^{N} \int_{0}^{t} d\tilde t \ \dot {\bv
  r}_{i}(\tilde t) \cdot {\bv u}_i$ in the dimensionless
position and time units.

\vspace{0.5cm}

\section{Fluctuation relation, {and convexity of $I(w)$} }
\label{appendix:fluctuationrelation}

\begin{figure}[h]
\begin{center}
\includegraphics[width=0.47\textwidth]{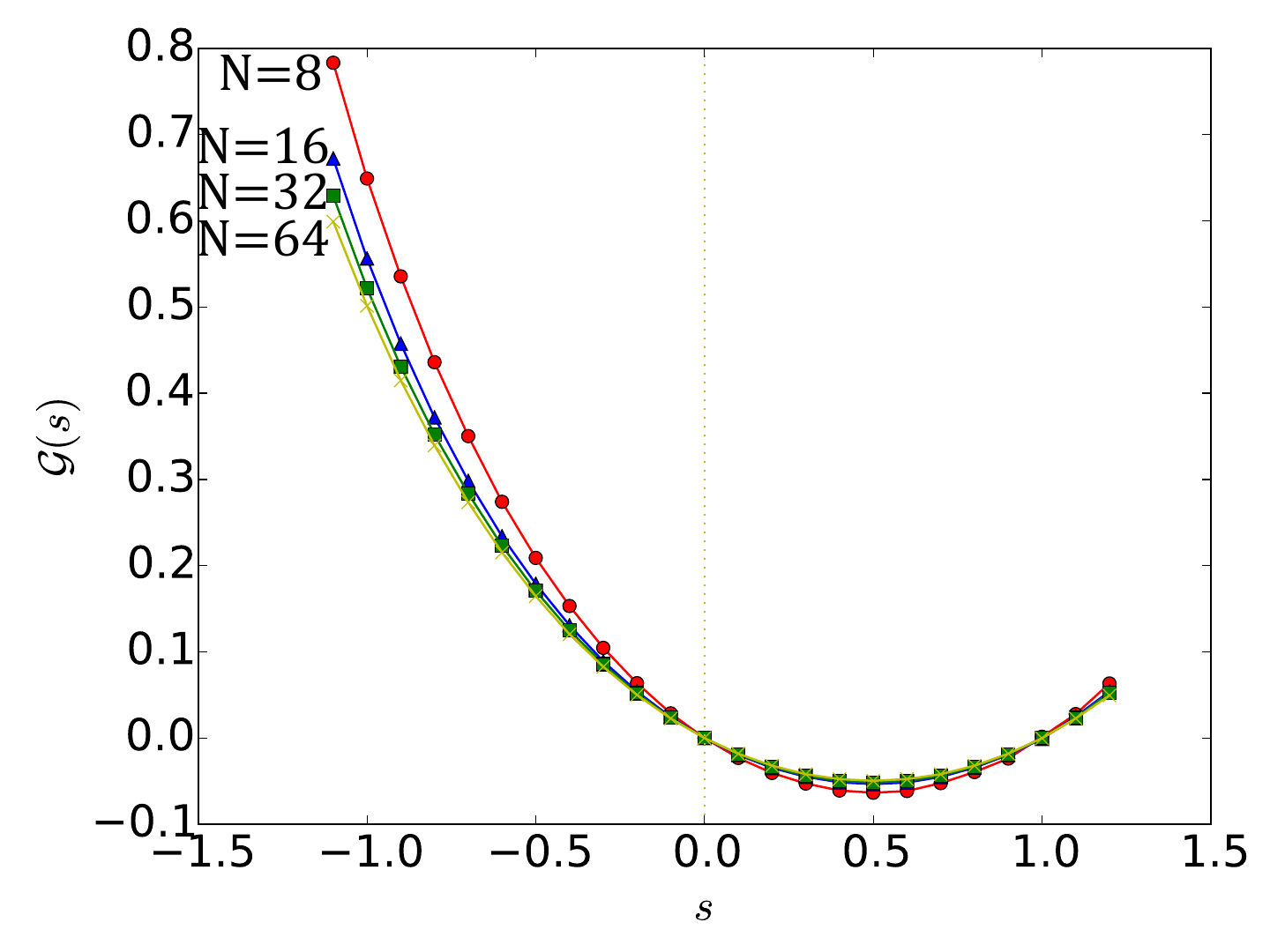} 
\caption{\label{Supp_fig:1} For ${\rm Pe}=1$ and $\phi=0.65$, the
  cumulant generating function ${\cal G}(s)$ is symmetric around
  $s_{\rm min}=1/2$, as predicted by Eq.~(\ref{supp_eq:1}). }
\end{center}
\end{figure}

The large deviation function of the active work satisfies the fluctuation theorem
\begin{equation}
{\cal G}(s) = {\cal G}({\rm Pe} - s).
\label{supp_eq:1}
\end{equation} 
This means that ${\cal G}(s)$ takes its minimum value at $s_{\rm
  min}={\rm Pe}/2$, where ${\cal G}'(s)$ vanishes. In
Fig.~\ref{Supp_fig:1}, we show a numerical example of ${\cal G}(s)$
for ${\rm Pe}=1$ which illustrates this symmetry property.

We now derive Eq.~\eqref{supp_eq:1}. First, let $P(\Omega)$ be the
probability density of a trajectory $\Omega=(\bv r_i(t'),
\theta_i(t'))$ with $t' \in [0,t]$.  We then define a time-reversed
trajectory as $\Omega^{T} = (\bv r_i(t - t'), \theta_i(t -
t'))$. Using standard methods~\cite{PhysRevLett.117.038103}, the ratio
between $P(\Omega)$ and $P(\Omega^T)$ can be computed as
\begin{equation}
\frac{P(\Omega)}{P(\Omega^T)} = \exp \left \{{\rm Pe} \  t N  w(t) - V(t) + V(0) \right \}
\end{equation}
where $V(t)$ is the total potential energy of the system at time $t$.
Multiplying both sides by $e^{-s t N w}P(\Omega^T)$ and summing upon all
possible $\Omega$, we get
\begin{equation}
\left \langle e^{-s t N w}  \right \rangle = \left \langle
e^{(s- {\rm Pe}) t N w - V(0) + V(t) } \right \rangle.
\end{equation}
The large time limit then immediately leads to 
the fluctuation theorem (\ref{supp_eq:1}).

An additional question within large-deviation theory 
is whether the limit in (\ref{eq:Gofs}) is finite, and whether the resulting $G(s)$ is analytic.  This discussion refers to finite systems, since it is clear from our results 
that the large-$N$ limit of ${\cal G}(s)$ can develop singularities.
If $G(s)$ is differentiable everywhere then $I(w)$ can be obtained from it by Legendre transformation and is also convex and continuous: all this
follows from the G\"artner-Ellis theorem~\cite{Touchette_LD}.  The finite-size scaling in our numerical work assumes that this theorem is applicable.

As
noted in Sec 3.2 of~\cite{Chetrite2015}, $G(s)$ is the largest eigenvalue of a differential operator (which is called the tilted generator); and if this operator
satisfies conditions for a  Perron-Frobenius theorem then it has a finite spectral gap. This is sufficient to establish that $G(s)$ is analytic, and hence the G\"artner-Ellis theorem.  The Perron-Frobenius theorem can be applied to systems such as \eqref{eqr_i} as long as the number of particles is finite, the domain in which they move is compact, the noise terms in \eqref{eqr_i} are additive (and non-zero), and the forces are bounded.

For the system considered here, there is one subtlety, which is that the interparticle forces $\bm{F}_{i,\rm ex}$ in \eqref{eqr_i} are not bounded. 
This prevents direct application of the Perron-Frobenius theorem.  From a mathematical point of view, this raises the possibility
that  large deviations might be realised by trajectories where two (or more) particles collapse onto the same point.   We do not have a mathematical 
proof that such trajectories can be neglected, but we see no physical reason why they would be relevant, and there is no evidence for them in our numerical computations. 
For this reason, we argue that the unbounded interaction forces in \eqref{eqr_i} can be truncated when particles come very close to the same point, without changing any of the behaviour that we find.  In such a modified system (with finite $N$), the Perron-Frobenius theorem applies, which means that $G(s)$ and hence $I(w)$ are both analytic, and are related by Legendre transformation

\section{Enhanced convergence of the cloning algorithm using modified dynamics}
\label{appendix:enhanced_convergence}

Our cloning algorithm gives access to the cumulant generating function
${\cal G}(s)$ in the limit of large number of clones. To enhance the
convergence of the algorithm, a generic strategy is to rely on
modified dynamics~\cite{Nemoto_Bouchet_Jack_Lecomte}. We now detail
the implementation of this strategy to sample the large deviations of
the active work in our model. We first introduce the following
modification of dynamics~\eqref{eqr_i}:
\begin{equation}
\dot {\bv r} _i = \alpha \frac{\sigma}{\ell_{\rm p}}\tilde { \bv F}_{i,\rm ex}  + (1 + f) {\bv u}_i(\theta_i) + \sqrt{\alpha \frac{2 \sigma}{\ell_{\rm p}}} \bv \eta _i,
\label{eqr_i_3}
\end{equation}
and 
\begin{equation}
\dot {\theta_i} =  \sqrt{\frac{2\sigma}{\ell_{\rm p}} } \xi_i.
\label{eqtheta_i_3}
\end{equation}
We denote $P_f(\Omega)$ the probability of $\Omega$ in this new
system. The following identity is then satisfied:
\begin{equation}
P(\Omega) e^{- s t N w} = P_f(\Omega) e^{- s t N \tilde w},
\end{equation}
where the new bias $\tilde w$ is defined as
\begin{equation}
\begin{split}
- s t N \tilde w  = & - \left ( s + \frac{f {\rm Pe}}{2} \right ) t N w  \\
& + \sum_{i=1}^{N} \int_0^{t} d\tilde t \left ( \frac{f {\rm Pe}}{2} + \frac{f^2 {\rm Pe}}{4} + \frac{f}{2} \tilde {\bv F}_{i,\rm ex} \cdot  {\bv u}_i \right ). 
\end{split}\label{eq:newbias}
\end{equation}
Simulating dynamics~\eqref{eqr_i_3} and~\eqref{eqtheta_i_3} with the
bias~\eqref{eq:newbias} is thus equivalent to
simulating~\eqref{eqr_i} with a bias $-stN w$. In practice we use
$f = - 2 s / {\rm Pe}$ so that the new bias $\tilde w$ reduces to
\begin{equation}
\tilde w =    1 -  \frac{s}{\rm Pe} + \frac{1}{t N \rm Pe}  \sum_{i=1}^{N} \int_{0}^{t} d\tilde t  \tilde {\bv F}_{i,\rm ex} \cdot  {\bv u}_i.
\end{equation}
which indeed produced faster convergence with the number of clones
used in the simulations.

\begin{figure}
\begin{center}
\includegraphics[width=0.42\textwidth]{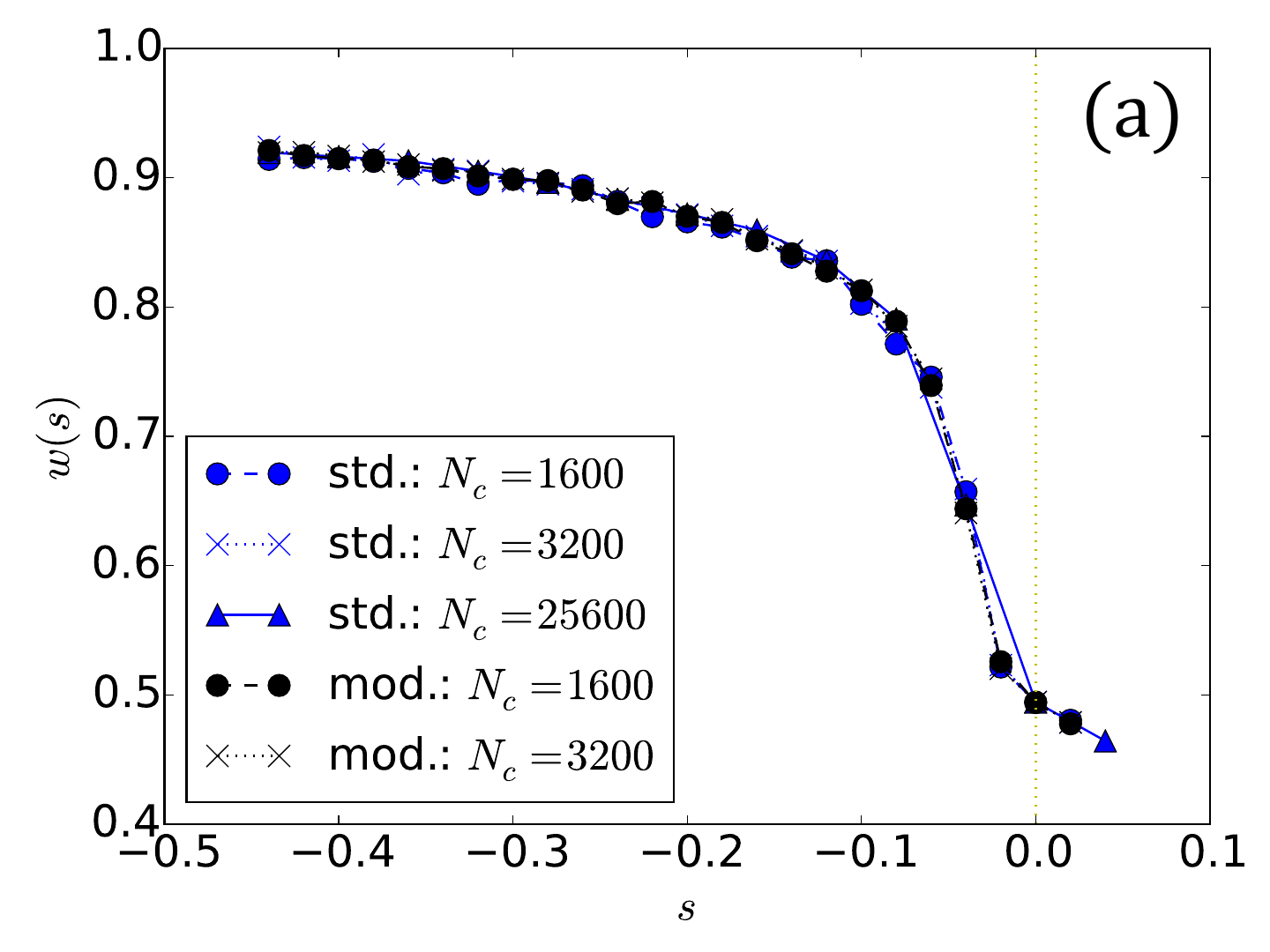} 
\includegraphics[width=0.42\textwidth]{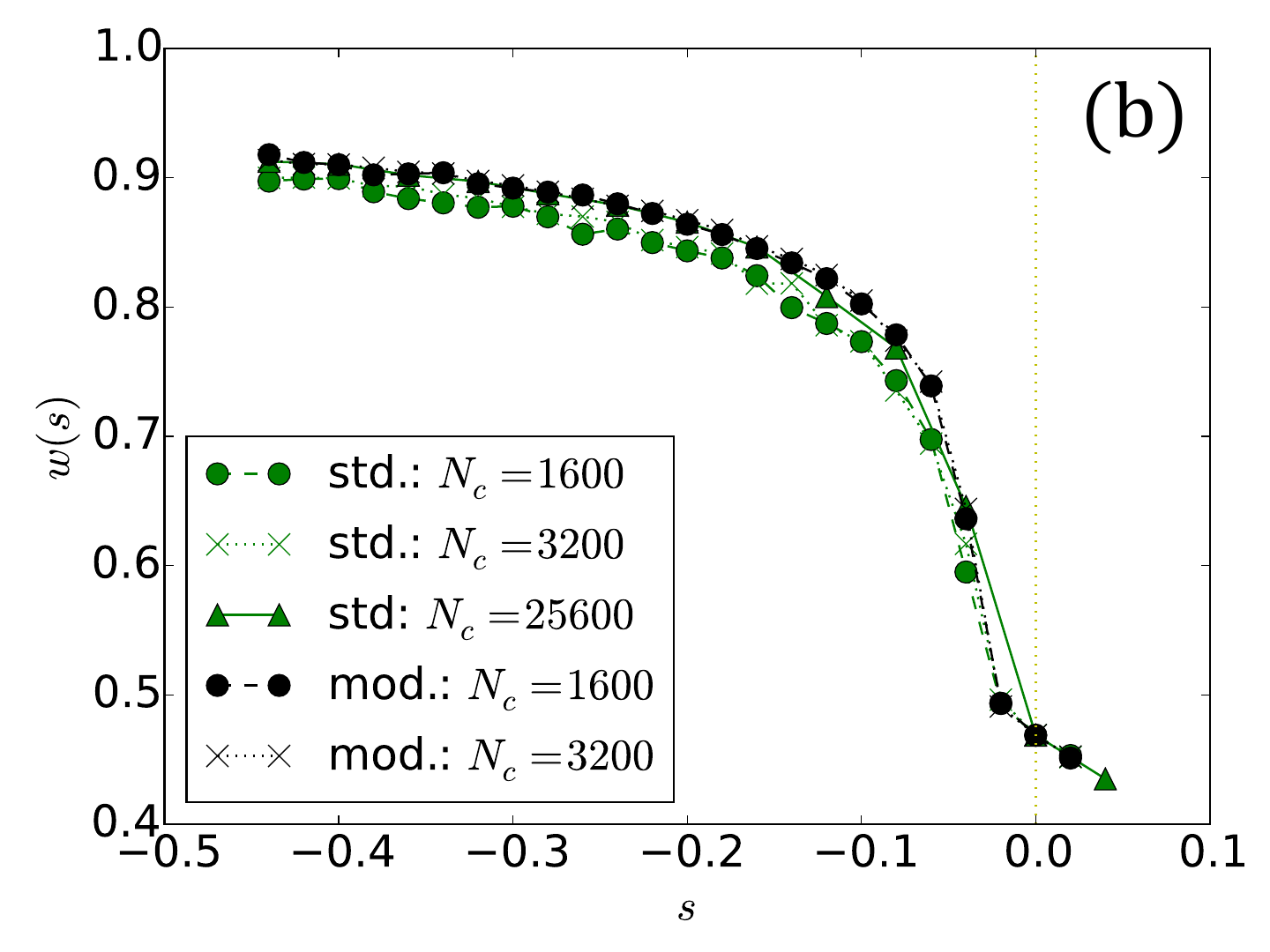} 
\includegraphics[width=0.42\textwidth]{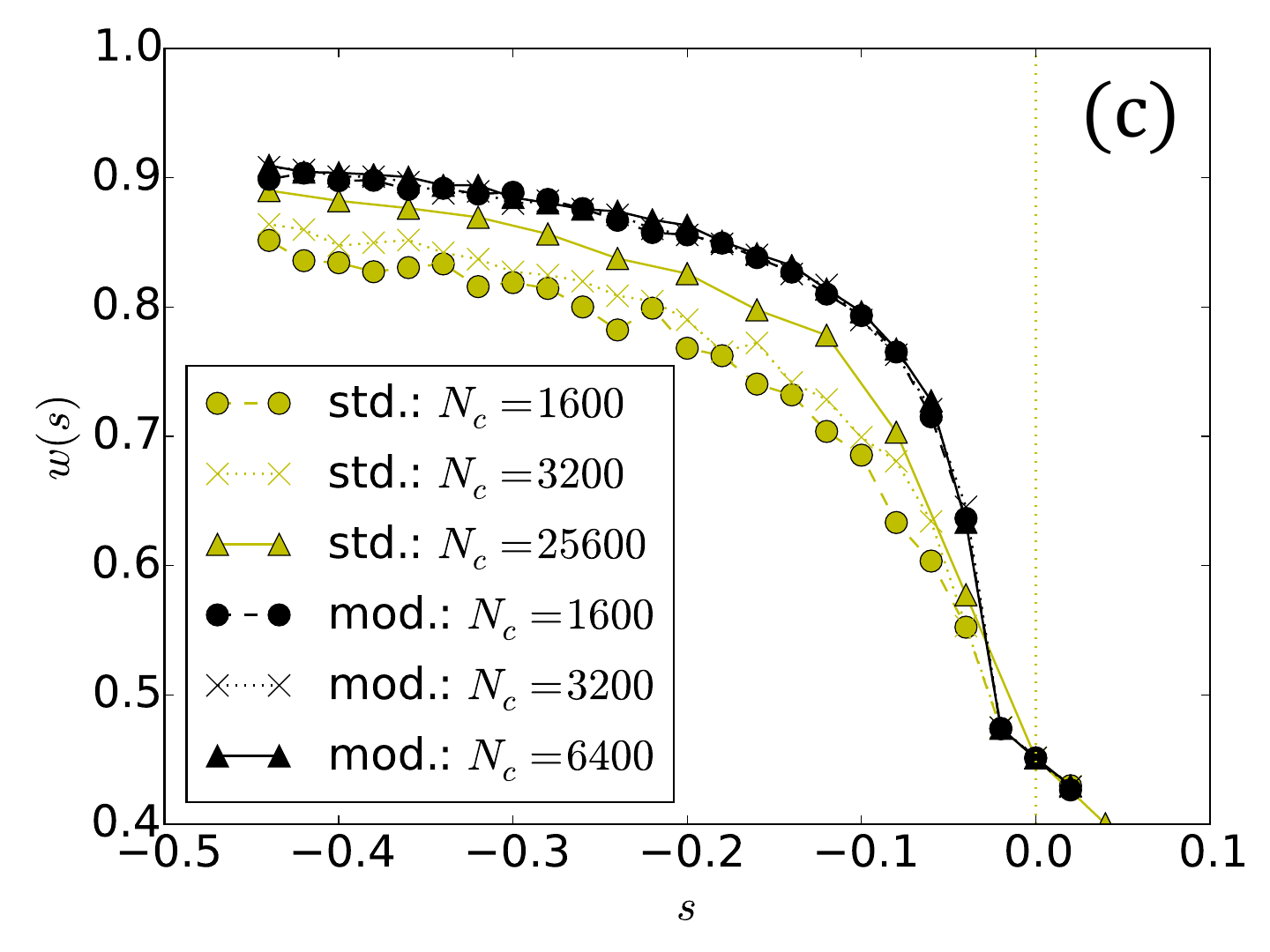} 
\caption{\label{fig:activework_comparison} The
  estimator of the active work obtained from the standard (std.) cloning
  algorithm (left-hand side of Eq.(\ref{eq:equality})) and modified (mod.)
  cloning algorithm (right-hand side of Eq.(\ref{eq:equality})) for $N=16$ (a), $N=32$ (b) and $N=64$ (c). The
  modified algorithm shows a much faster convergence as the number of
  clones is increased. Its limiting value agrees with the standard
  algorithm when the latter has converged.}
\end{center}
\end{figure}

{To characterize the CM state, we add another modifying force described
as follows:
\begin{equation}
\label{eqr_i_4}
\dot {\bv r} _i = \frac{1}{{\rm Pe}} \bv F_{i,\rm ex} + \left ( 1 -\frac{2s}{\rm Pe} \right ) {\bv u}(\theta_i) + \sqrt{\frac{2}{\rm Pe} } 
\bv \eta_i ,
\end{equation}
and
\begin{equation}
\dot {\theta_i}  = - g N \frac{\partial}{\partial \theta_i}\hat{\nu}^2 + \sqrt{\frac{2}{ \alpha {\rm Pe}}} \xi_i,
\end{equation}
where $g$ is a parameter whose value is discussed later.  Similarly,
the probability of the trajectory $\omega$ in this modified system,
$P_{\rm mod}$, is given by
\begin{equation}
\begin{split}
& P_{\rm mod}(\omega) \propto  \exp \Bigg \{  -\frac{\rm Pe}{4} \int dt \sum_i \bigg (\dot {\bv r} _i - \frac{1}{\rm Pe}\bv F_{i,\rm ex} -  \\
& \left ( 1 -\frac{2s}{\rm Pe} \right ){\bv u}(\theta_i) \bigg )^2 -\frac{\alpha \rm Pe}{4} \int dt \sum_i \left ( \dot \theta_i + g N \frac{\partial}{\partial \theta_i} \hat{\nu}^2  \right )^2  \\
& - \frac{1}{4 \rm Pe} \int dt \sum_i\frac{\partial}{\partial {\bv r} _i}\bv F_{i,\rm ex}
+ \frac{1}{2} \int_{0}^{\tau} dt \sum_i g N\frac{\partial^2 \hat{\nu}^2 }{\partial \theta_i^2}  \bigg \}. 
\end{split}
\end{equation}
By taking the ratio between $P(\omega)$ and $P_{\rm mod}(\omega)$, we get
\begin{equation}
\begin{split}
& \frac{P_{\rm mod}(\omega)}{P(\omega)} \simeq \exp \Bigg \{  - s \tau N w + \int_0^{\tau} dt \bigg [s N - \frac{s^2}{\rm Pe}N \\
&+ \frac{s}{\rm Pe} \sum_i \bv u_i \cdot \bv F_{i,\rm ex}   + g - g N \hat{\nu}^2 - \frac{g^2 \alpha \rm Pe}{\hat{\nu}^2}\sum_{i}\sin^2(\theta_i-\varphi) \bigg ]   \Bigg \},
\end{split}
\end{equation}
where $w$ is the active work introduced in the main text. 
By defining the modified active work $w_{\rm mod}$ as
\begin{equation}
\begin{split}
& w_{\rm mod} = \frac{1}{\tau}\int_0^{\tau} dt \bigg [1 - \frac{s}{\rm Pe} + \frac{1}{N \rm Pe} \sum_i \bv u_i \cdot \bv F_{i,\rm ex}   + \frac{g}{sN} \\
& - \frac{g}{s} \hat{\nu}^2 - \frac{g^2 \alpha \rm Pe}{s N \hat{\nu}^2}\sum_{i}\sin^2(\theta_i-\varphi) \bigg ] 
\end{split}
\end{equation}
we thus get
\begin{equation}\label{eq:equality}
P(\omega) e^{-s\tau N w} = P_{\rm mod}(\omega) e^{-s \tau N w_{\rm mod}}.
\end{equation}
Note that $g$ is a free parameter here: the
equality~\eqref{eq:equality} holds irrespective of the value of
$g$.

As discussed in~\cite{Nemoto_Bouchet_Jack_Lecomte} and in Appendix \ref{Appendix:optimal_variational},
there is an optimal modification to the dynamics -- if this could be
found, then the cloning algorithm would have zero error and $w_{\rm
  mod}$ in~\eqref{eq:equality} would become a simple number
(independent of the trajectory), equal to $-G(s)/s$.  However, finding
the optimal modification is as difficult as solving the
large-deviation problem analytically, and is out of reach for most
problems, including this one. Hence, the modifying forces used here
are not optimal in the sense of~\cite{Nemoto_Bouchet_Jack_Lecomte} but
we may still choose $g$ so as to enfore the following equality:
\begin{equation}
\left \langle -s w_{\rm mod} \right \rangle_{\rm mod}  = \hat G(s)
\label{eq:gcondition},
\end{equation}
where $ \left \langle \ \right \rangle_{\rm mod}$ means the average in
the modified dynamics (obtained from the cloning algorithm) and $\hat
G(s)$ is the estimator of the cumulant generating function within the
cloning algorithm. We found that this is an efficient way to choose
our modifying force.

In Fig.~\ref{fig:activework_comparison}, we compare the standard
cloning algorithm (left-hand side of Eq.(\ref{eq:gcondition})) and the
modified one (right-hand side of Eq.(\ref{eq:gcondition})) by plotting
the active work as a function of s. We see that for $N=16$ and $N=32$,
both algorithm lead to the same function $w(s)$, but that the modified
dynamics converges much faster as the number of clones $N_c$
increases. For $N=64$, however, the standard algorithm shows
a very slow convergence, unlike our modified algorithm.}

\section{The parameters used for Population Monte-Carlo method}

Here we summarize the parameters used to get the results in the main text. Convergence is obtained using the number of clones
$N_c=25600$. The time-step of the simulations is $dt=0.001$; cloning
steps are performed each $\Delta t = 0.01$. The simulation length
varies from $t=30000$ to $300000$, depending on the values of $N$ and
$s$. We checked the convergence with respect to the cloning parameters
$N_c$, $\delta t$, $t$, for all values of $s$ except for the immediate
vicinity of the PSA transition point ($s>0$). Around this transition
point, we observe a slight unphysical concavity of ${\cal G}(s)$ which
is a signature that perfect convergence is out of reach of our
simulations (see Fig.~\ref{Supp_fig:sc_scaling}(a)).

\section{Derivation of the polarization dynamics}
\label{appendix:derivation}

This appendix is devoted to the derivation of the dynamics of the
stochastic polarization $\hat\nu$ defined in Eq.~(\ref{eq:defnu}) of
the main text. It can be written as
\begin{equation}\label{eq:pol}
	\hat\nu = \frac{1}{N} \sqrt{ \sum_{\{i,j\}=1}^N \cos(\theta_i-\theta_j) } .
\end{equation}
We introduce a global phase $\varphi$ such as
\begin{equation}\label{eq:phi}
	\hat\nu e^{\varphi} \equiv \frac{1}{N} \sum_{j=1}^N e^{\text{i}\theta_j} .
\end{equation}
Using It\^o's lemma, taking time derivative of Eq.~\eqref{eq:pol} gives
\begin{equation}\label{eq:pol_dyn}
	\begin{aligned}
		\dot{\hat\nu} &= - \frac{\sqrt{2D_\text{r}}}{N^2\hat\nu} \sum_{\{i,j\}=1}^N \xi_i\sin(\theta_i-\theta_j)
        \\
        &\quad + \frac{D_\text{r}}{N} \sum_{k=1}^N \frac{d^2}{d \theta_k^2} \sqrt{\sum_{\{i,j\}=1}^N \cos(\theta_i-\theta_j)} ,
	\end{aligned}
\end{equation}
where we have used $\dot\theta_i=\sqrt{2D_\text{r}}\xi_i$, and $\xi_i$ is a zero-mean unit-variance Gaussian white noise. The second term can be written as
\begin{widetext}
  \begin{equation}\label{eq:ito}
      \begin{aligned}
          \sum_{k=1}^N \frac{d^2}{d \theta_k^2} \sqrt{\sum_{\{i,j\}=1}^N \cos(\theta_i-\theta_j)} &= - \sum_{\{k,l\}=1}^N \frac{d}{d \theta_k} \frac{\sin(\theta_k-\theta_l)}{\sqrt{\sum_{i,j} \cos(\theta_i-\theta_j)}}
          \\
          &= \frac{N-\sum_{k,l}\cos(\theta_k-\theta_l)}{\sqrt{\sum_{i,j} \cos(\theta_i-\theta_j)}} - \sum_{\{k,l,m\}=1}^N \frac{\sin(\theta_k-\theta_l)\sin(\theta_k-\theta_m)}{\big[\sum_{i,j} \cos(\theta_i-\theta_j)\big]^{3/2}}
          \\
          &= \frac{1}{\hat\nu} - N \hat\nu - \frac{1}{(N\hat\nu)^3}\sum_{\{k,l,m\}=1}^N \sin(\theta_k-\theta_l)\sin(\theta_k-\theta_m) .
      \end{aligned}
  \end{equation}
\end{widetext}
The noise term appearing in~\eqref{eq:pol_dyn} is denoted by
\begin{equation}
	\Lambda \equiv \sum_{\{i,j\}=1}^N\xi_i\sin(\theta_i-\theta_j) .
\end{equation}
It is a zero-mean Gaussian white noise with correlations
\begin{equation}\label{eq:lambda}
	\begin{aligned}
		\langle\Lambda(t)\Lambda(0)\rangle &= \sum_{\{i,j,k,l\}=1}^N \langle\xi_i(t)\xi_k(0)\rangle \sin[\theta_i(t)-\theta_j(t)]
        \\
        &\qquad\qquad \times\sin[\theta_k(0)-\theta_l(0)]
        \\
        &= \delta(t) \sum_{\{j,k,l\}=1}^N \sin(\theta_k-\theta_j)\sin(\theta_k-\theta_l) .
	\end{aligned}
\end{equation}
To proceed further, we note that the sum can be simplified, using~\eqref{eq:phi}, as
\begin{equation}
	\begin{aligned}
      \sum_{\{j,k,l\}=1}^N &\sin(\theta_k-\theta_j)\sin(\theta_k-\theta_l)
      \\
      &= (N\hat\nu)^2 \sum_{i=1}^N \sin(\theta_i-\varphi)^2
      \\
      &= (N\hat\nu)^2 \int_0^{2\pi} \sin(\theta-\varphi)^2 \psi(\theta,t) d\theta ,
    \end{aligned}
\end{equation}
where we have introduced the angular distribution
$\psi(\theta,t)\equiv\sum_i\delta[\theta-\theta_i(t)]$. We now assume that
$\psi$ is close to uniform, which should hold for $\hat\nu\ll1$ and
large $N$, to obtain
\begin{equation}
	\begin{aligned}
      &\sum_{\{j,k,l\}=1}^N \sin(\theta_k-\theta_j)\sin(\theta_k-\theta_l)
      \\
      &\quad
      \xrightarrow[\hat \nu \ll1,~N\gg1]{}
      \frac{N^3\hat\nu^2}{2\pi} \int_0^{2\pi} \sin(\theta-\varphi)^2 d\theta = \frac{N^3\hat\nu^2}{2} .
    \end{aligned}
\end{equation}
{Finally, substituting this result in~\eqref{eq:ito} and~\eqref{eq:lambda}, then \eqref{eq:pol_dyn} reduces to
\begin{equation}
\frac{d\hat \nu}{dt} = D_{\rm r} \left [\frac{1}{2N\hat \nu} - \hat \nu \right ] + \sqrt{\frac{D_{\rm r}}{N}} \xi .
\label{eq:p}
\end{equation}
which is a closed (autonomous) equation for the evolution of $\hat\nu$.}

\section{CGF of the time-averaged total orientation $\bar \nu$}
\label{appendix:cgf}

In this appendix, we consider the cumulant generating function of
$\bar \nu$, defined as
\begin{equation}
  {\cal H}(\varsigma)=\frac 1 N \lim_{t\rightarrow \infty} \frac 1 t \log
\left \langle e^{- t N \varsigma \bar \nu}\right \rangle\; .
\end{equation}
We work under the assumption that $\hat \nu \ll 1$, $N\rightarrow
\infty$ so that we can use the time-evolution equation for $\hat \nu$
given by (\ref{eq:p}). We introduce the rescaled variable
$q=\hat\nu\sqrt{N}$, whose dynamics is given by
\begin{equation}\label{eq:dynq}
\frac{dq}{dt} = D_{\rm r} \left [\frac{1}{2q} - q \right ] + \sqrt{D_{\rm r}} \xi\;.
\end{equation}
Note that~\eqref{eq:dynq} is {\it independent} of $N$.  We then
consider the cumulant generating function of the time-averaged value
of $q$:
\begin{equation}
f(k) =  \lim_{t \rightarrow \infty}\frac{1}{t} \log \left \langle e^{-k\int_{0}^t d\tilde t q(\tilde t)}  \right \rangle\;.
\end{equation}
The CGF $f(k)$ is the largest eigenvalue of the following operator:
\begin{equation}
L_{k}[\cdot] = - \frac{\partial}{\partial q}\left [D_r \left (\frac{1}{2q} - q \right )  \cdot \right ] + \frac{D_{\rm r}}{2} \frac{\partial^2}{\partial q^2} \left [ \cdot \right ]
- k q.
\label{eq:operator}
\end{equation}
Since $L_{k}$ is independent of $N$, $f(k)$ is a well-defined smooth
function in the $N\rightarrow \infty$ limit:
\begin{equation}
f(k) =  \lim_{N,t \rightarrow \infty}\frac{1}{t} \log \left \langle e^{-k\int d\tilde t q(\tilde t)}  \right \rangle.
\end{equation}

\begin{figure}
\begin{center}
\includegraphics[width=0.43\textwidth]{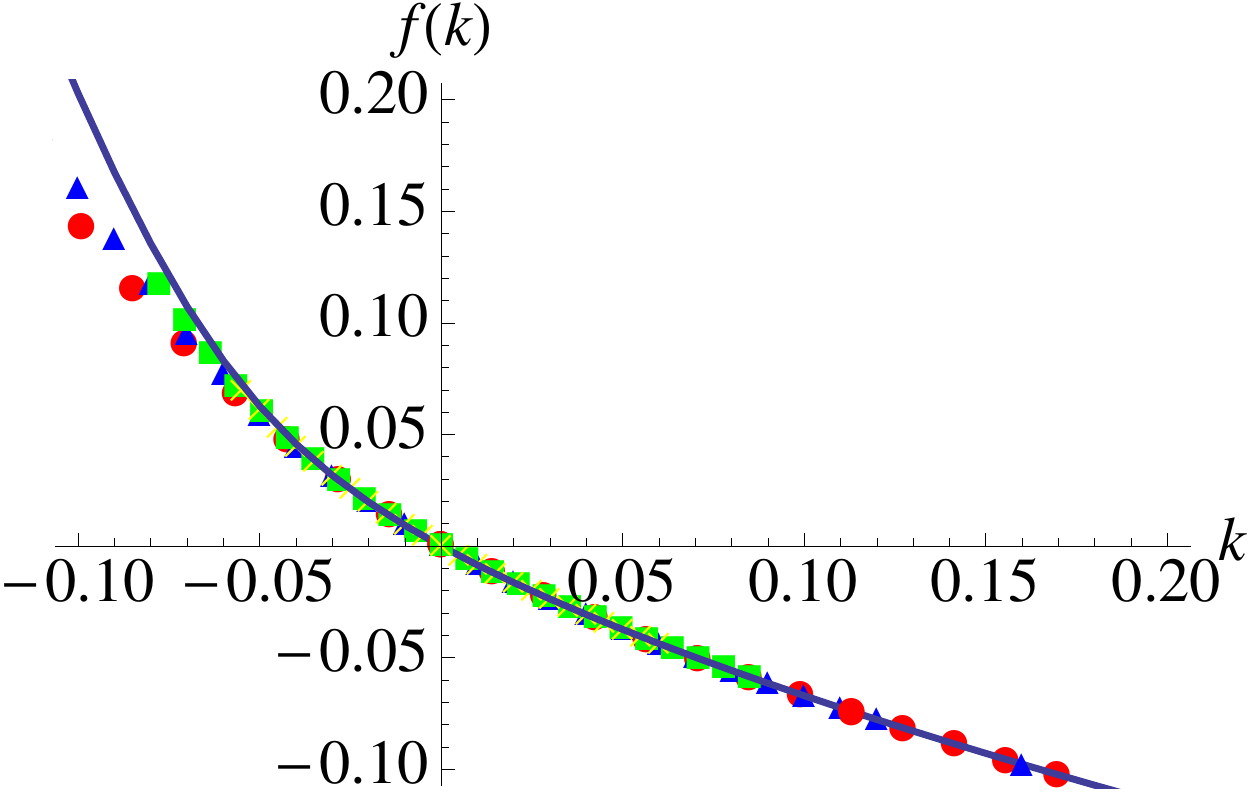}
\caption{\label{fig:2} An example of $f(k)$ obtained from numerical diagonalization of the operator (\ref{eq:operator}) in blue solid line. We set $D_{\rm r} = 1/40$. We also plot $N{\cal H}(k/\sqrt{N})$ obtained from the application of the cloning algorithm to $N$ active rotors. The data points for $N=8,16,32,64$ correspond to red-circle, blue-triangle green-square and yellow cross. We can see clear overlap of the data points to a single line, demonstrating the validity of (\ref{f(k)_H(k)}).} 
\end{center}
\end{figure}

Using $k=\varsigma \sqrt{N}$, ${\cal H}(\varsigma)$  can now be expressed as
\begin{equation}
{\cal H}(\varsigma) = \frac{f(\varsigma\sqrt{N})}{N},
\label{H(h)_f(h)}
\end{equation}
or, conversely,
\begin{equation}
f(k) =  N {\cal H}\left (\frac{k}{\sqrt{N}} \right ).
\label{f(k)_H(k)}
\end{equation}
In Fig.~\ref{fig:2}, we numerically demonstrate~\eqref{f(k)_H(k)}. To
do so, we compare the results obtained by applying the cloning
algorithm to the dynamics~\eqref{eqtheta_i_2} of $N$ independent
rotors, which yields the right-hand side of (\ref{f(k)_H(k)})), with
the result of the numerical diagonalization of the operator
(\ref{eq:operator}), which yields the left-hand side of
(\ref{f(k)_H(k)}).  The results of the cloning algorithm
for several $N$ clearly collapse onto a single function $f(k)$. Note
that this overlap is satisfied not only for positive $s$ (where the
assumption $\hat \nu \ll 1$ is safely satisfied) but also for negative
$s$ close to the origin.

\section{Finite-size scaling to estimate PSA transition point in $N\rightarrow \infty$}
\label{appendix:Finite-size}

\begin{figure}
\begin{center}
\includegraphics[width=0.45\textwidth]{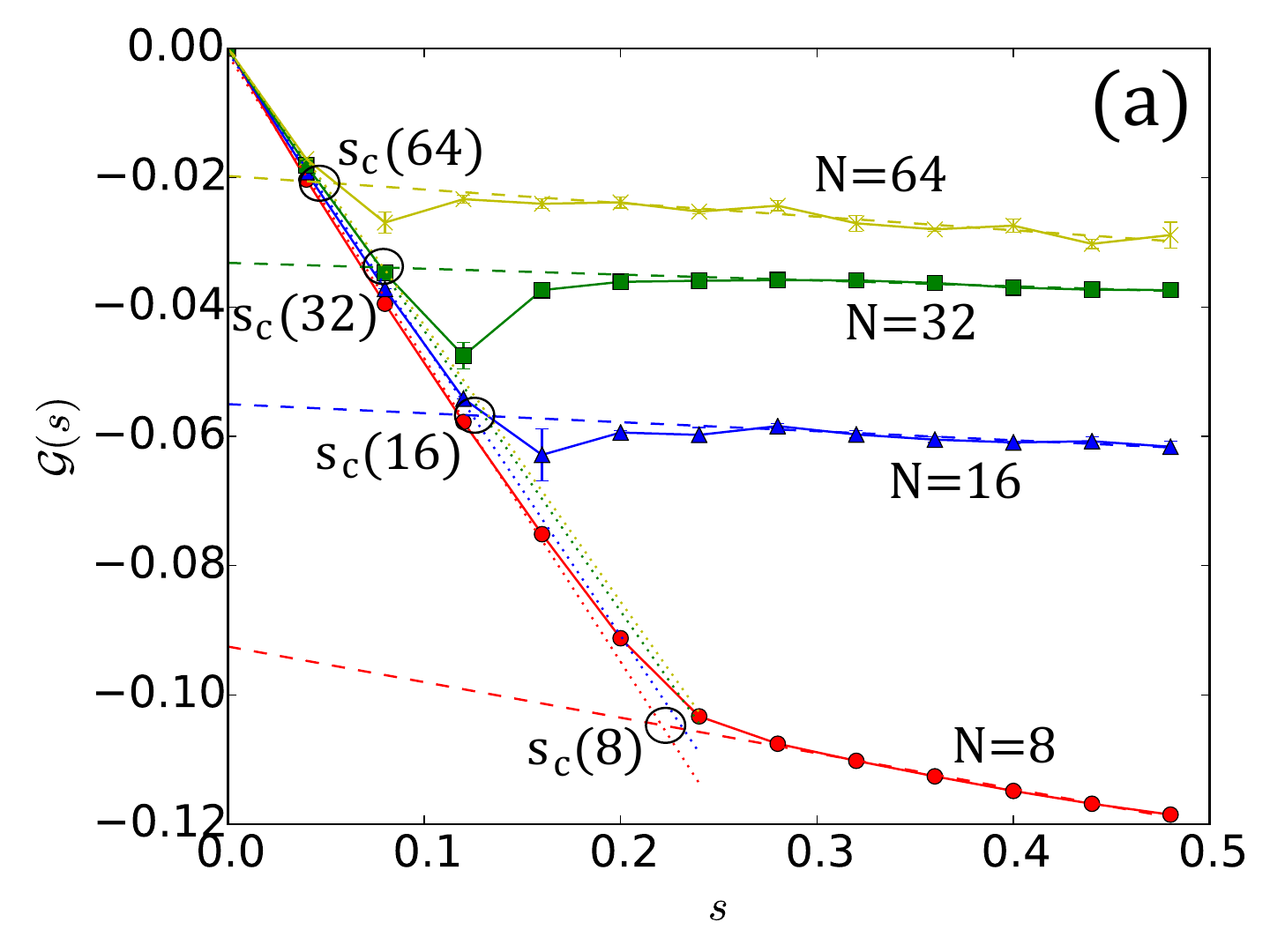} 
\includegraphics[width=0.45\textwidth]{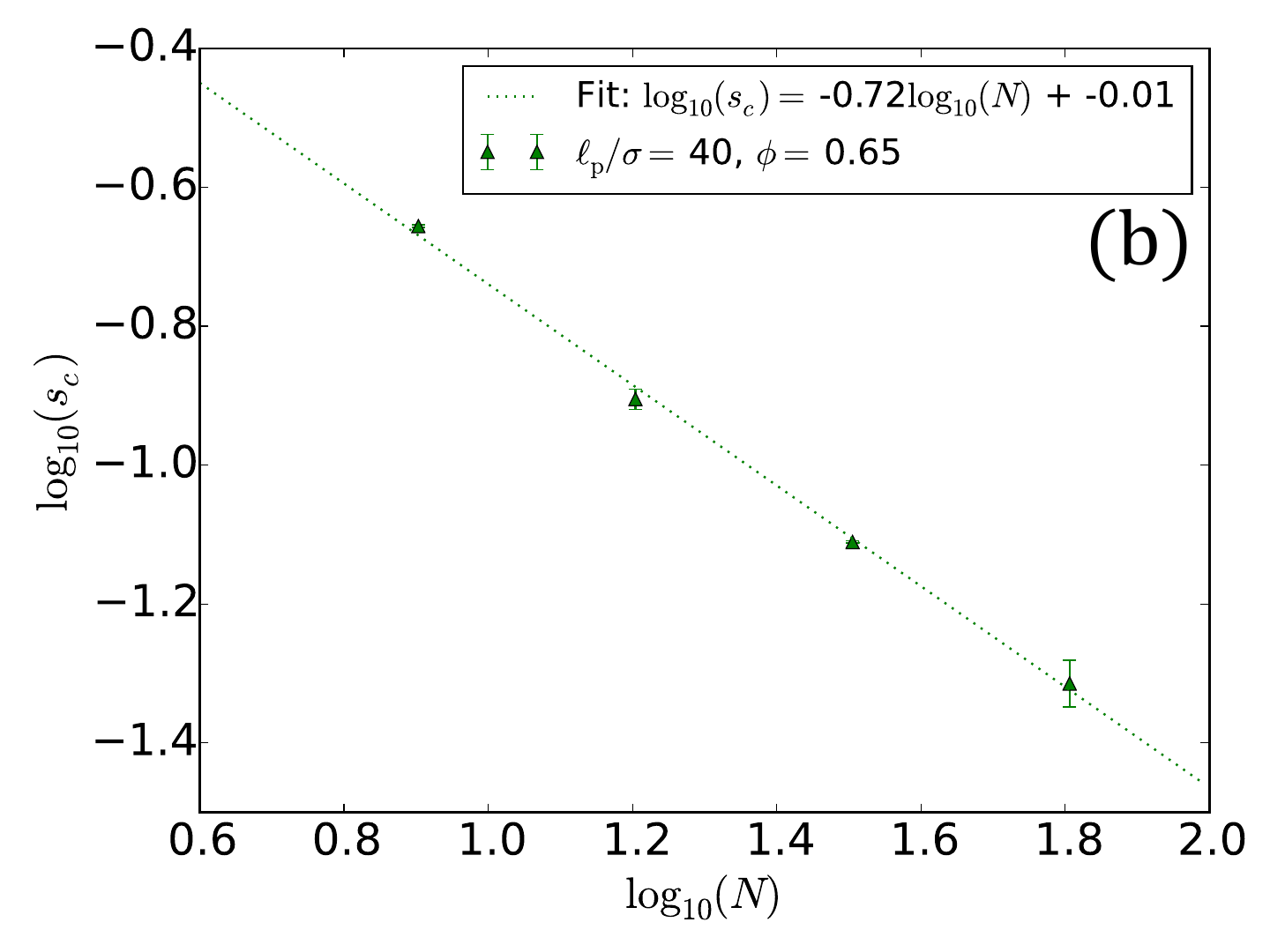}
\caption{\label{Supp_fig:sc_scaling} {\bf (a)} How to estimate
  $s_c(N)$ from ${\cal G}(s)$.  For $s$ slightly larger than $s_c$,
  the required number of clones $N_c$ needed to observe the
  convergence of the cloning algorithm rapidly increases beyond what
  can be reached numerically. As a consequence, we see an artificial
  violation of the concavity of ${\cal G}(s)$.  To interpolate the
  correct shape of ${\cal G}(s)$, and locate $s_c(N)$, we use the data
  where the concavity is not violated. In both panels, dashed and
  dotted straight lines are obtained by fitting the data in inactive
  ($s \gg s_c(N)$) and active ($s\ll s_c(N)$) regions,
  respectively. Assuming a sharp kink~\cite{Nemoto_Jack_Lecomte}, we
  obtain the estimated value of $s_c(N)$ as the crossing point of
  these two lines (indicated as the black circles for each $N$).  {\bf
    (b)} $s_c(N)$ estimated from the finite-size scaling on the data
  up to $N=64$ (for $\ell_{\rm p} = 40\sigma$, $\phi=0.65$) shows a
  power law decay with respect to $N$: $s_c(N) \sim N^{-a}$ with a
  positive constant $a$. In particular, this implies
  $\lim_{N\rightarrow \infty}s_c(N)=0$.  }
\end{center}
\end{figure}

We denote by $s_c(N)$ the PSA transition point for finite system size
$N$. It is defined as the value of $s$ that maximizes the second
derivative of ${\cal G}(s)$ for positive $s$. The obstacle to estimate
$s_c(N)$ is that there are strong finite-size effects with respect to
the number of clones $N_c$ around $s_c(N)$ (that artificially violate
the convexity of ${\cal G}(s)$ as seen in
Fig.~\ref{Supp_fig:sc_scaling}(a)). To overcome this difficulty, we
extract $s_c(N)$ from the crossing point of the straight lines
obtained by fitting the data for $s<s_c(N)$ and $s_c(N)<s$, as shown
in Fig.~\ref{Supp_fig:sc_scaling}(a). Due to the convexity of $G(s)$,
the crossing point determined in this way gives a good approximation
of $s_c(N)$~\cite{Nemoto_Jack_Lecomte}. We then plot $s_c(N)$ as a
function of $N$ and extrapolate $\lim_{N\rightarrow \infty}s_c(N)$. As
seen from Fig.~\ref{Supp_fig:sc_scaling}(b), $s_c(N)$ is consistent
with a convergence to zero, with a power law: $s_c(N) \sim N^{-a}$.

\section{Optimal control argument for dynamical arrest of a MIPS cluster (PSA transition)}
\label{Appendix:optimal_variational}

We consider a system that is obtained by adding additional ``control'' forces to \eqref{eqr_i}, leading to
\begin{align}
\dot {\bv r} _i &= B^r_i({\bv r},\theta) + \mu \bv F_{i,\rm ex} + \vp {\bv u}(\theta_i) + \sqrt{2 D} 
\bv \eta_i ,\nonumber \\ 
\dot {\theta_i} &  =  B^\theta_i({\bv r},\theta) + \sqrt{2 D_{\rm r}} \xi_i.
\label{control}
\end{align}
The control forces $B^r,B^\theta$ depend on the co-ordinates of all particles.
A general result in large-deviation theory (see, e.g., Eq.~(54) of~\cite{TouchetteJSM2015} as well as~\cite{Nemoto2011} and the discussion in Sec.~4 of~\cite{JackSollichEPJ2015}) is that
\begin{equation}
I(w) = \inf_{B^r,B^\theta \colon \langle w \rangle_{\rm control}=w} \Phi(B^r,B^\theta),
\label{equ:inf}
\end{equation}
where $\Phi$ is a ``cost function'', and the infimum runs over those forces for which the  steady state of (\ref{control}) has a mean active work $w$.  The cost function $\Phi$ is the relative entropy between two ensembles of trajectories, which are the (unbiased) steady state of the original system, and the steady state of the controlled system.  This relative entropy is related to the large-deviation rate function at level-2.5: in the present context it is simply~\cite[Eq.~(76)]{TouchetteJSM2015} 
\begin{equation}
\Phi(B^r,B^\theta) = \left\langle \sum_i \left[ \frac{(B_i^r)^2}{4D} + \frac{(B_i^\theta)^2}{4D_r}\right] \right\rangle_{\rm control},
\label{equ:Phi}
\end{equation}
where the average is taken in the steady state of (\ref{control}).

From (\ref{equ:inf}) one has that for any control forces $B$ that
realise the desired active work, then $I(w)\leq\Phi(B^r,B^\theta)$.
To establish the existence of a phase transition at $s=0$, it is
sufficient to find (for each $N$) some $B$ that realise the desired
active work, with $\lim_{N\to\infty}N^{-1}\Phi(B^r,B^\theta) = 0$
(that is, $\Phi$ is subextensive).  In this case $ {\cal
  I}_{\infty}(w) \equiv \lim_{N\to\infty}N^{-1} I(w) \leq
\lim_{N\to\infty}N^{-1}\Phi(B^r,B^\theta) = 0$.  The rate function is
non-negative so this is sufficient to show that $ {\cal
  I}_{\infty}(w)=0$.

\begin{figure}
\begin{center}
\includegraphics[width=0.23\textwidth]{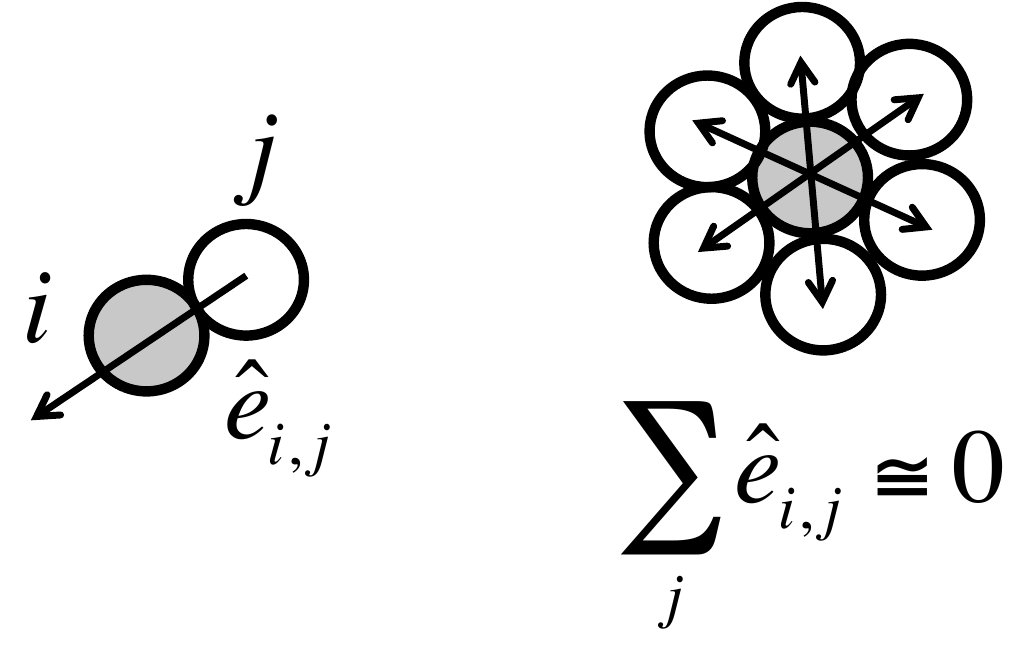} 
\caption{\label{fig:symmetry}
{Schematic figure to show $\sum_j\hat{\bv e}_{i,j}\simeq 0$ when a particle (indicated as gray color) is surrounded by six particles.}}
\end{center}
\end{figure}

\begin{figure}
\begin{center}
\includegraphics[totalheight=0.22\textwidth]{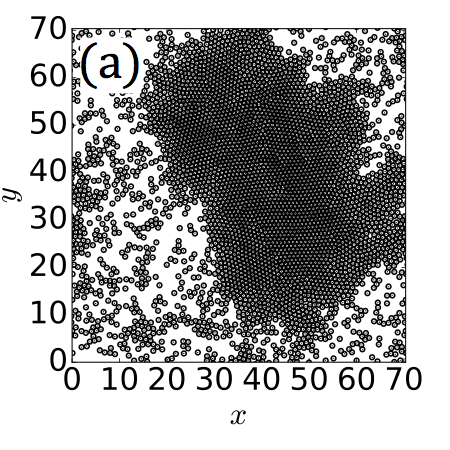} 
\hspace{-.4cm}
\includegraphics[totalheight=0.22\textwidth]{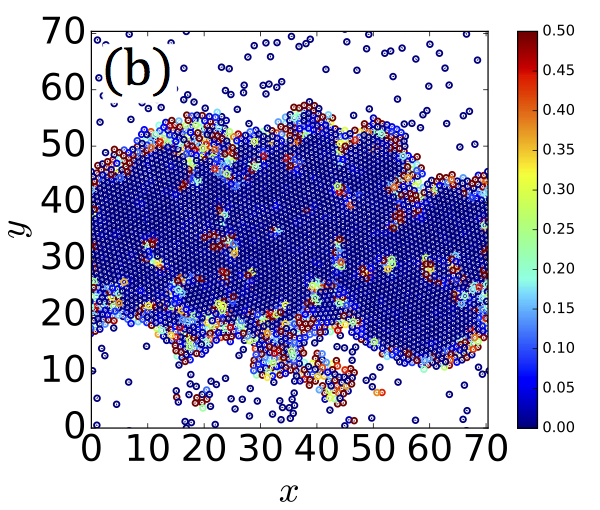} \\
\hspace{-0.5cm}
\includegraphics[width=0.42\textwidth]{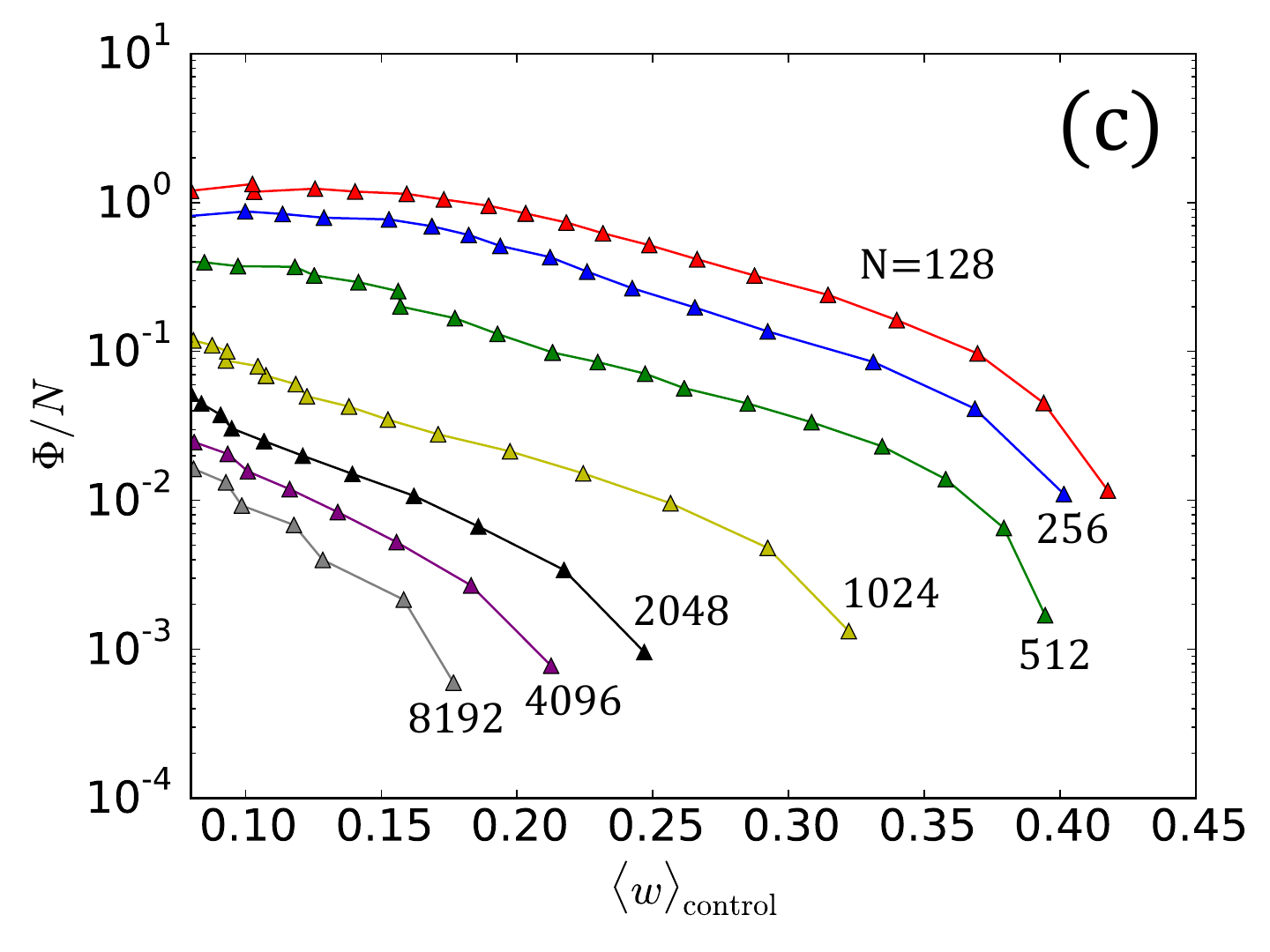} 
\caption{\label{fig:clustering} { {\bf (a)} Snapshot of the original
    dynamics, for $N=4096$ particles, leading to $\langle w
    \rangle \simeq 0.25$.  {\bf (b)} Snapshot of the dynamics with the
    control torque \eqref{eq:auxB} and $g=0.22$. The control torques
    clearly reduce the number of particles in the gas phase, leading to
    a lower active work $\langle w\rangle \simeq 0.05$. The color code
    corresponds to the contribution of
    each particle to the cost function~\eqref{equ:Phi}, normalized by
    $g^2$. The cost is clearly dominated by the subextensive
    contribution of the particles localized at the boundary of the
    main cluster. 
    {\bf (c)} The upper bound $\Phi$ of the large
    deviation function as a function of $\langle w \rangle_g = \left
    \langle w \right \rangle_{\rm control}$, normalized by the number
    of particles. The bound decreases as the system size
    increases.}}
\end{center}
\end{figure}

To illustrate how this argument works, recall the case of kinetically constrained models.  In that case the variational principle (\ref{equ:inf}) simplifies~\cite{JackSollichEPJ2015}, because the controlled systems are at equilibrium and are fully characterised by their Boltzmann distributions. One may then find $B$ such that the system is localised in a single state and the corresponding $\Phi$ is the escape rate from that site.  In KCMs there are configurations for which this escape rate is subextensive, leading to ${\cal I}_{\infty}(a)=0$ for $0<a<\langle a \rangle$, (where $a$ is the dynamical activity).

In the present context, we suppose that the natural state of the system is phase-separated (due to MIPS) and we consider a controlled steady state that is also phase-separated.  We take $B^r=0$, and we apply torques $B^\theta$ that act on the particles near the boundary of the dense cluster, which favour orientations pointing towards the cluster.  These torques will help to reinforce the MIPS state, and they also act to compress the cluster, so that its density will increase, which tends to reduce particle motion.  For any control force of this type, the only terms which contribute in (\ref{equ:Phi}) are from particles on the cluster boundary, so the number of such terms is subextensive.  Hence $\Phi$ is subextensive (assuming that the $B^\theta_i$ are bounded).  This means that $ {\cal I}_{\infty}(w)=0$ for any value of the active work that can be realised by a perturbation of this type.  Our data (Fig.~\ref{fig:RateFunction}) indicate that values of $w$ close to zero can be achieved with a subextensive cost, just as happens in KCMs. 

{Building such control forces and torques explicitly, that would apply
only to particles located at the boundary of the cluster, is a
numerical challenge. We can nevertheless test our hypothesis by
considering the following protocol
\begin{equation}
B_i^\theta = - g \frac{\partial}{\partial \theta_i} \sum_j \hat{\bv{e}}_{i,j} \cdot \bv u_i(\theta_i)
\label{eq:auxB}
\end{equation}
(with $B_i^r=0$). Here, $g>0$ is a constant parameter and $\hat{\bv
  e}_{i,j}$ is a unit vector from the particle $j$ to the particle $i$
when they interact and zero otherwise: $\hat{\bv e}_{i,j}
=\Theta(2^{1/6}- r_{i,j}) \bv r_{i,j}/r_{i,j}$ with $\bv r_{i,j}\equiv
\bv r_i - \bv r_j$. The torques~\eqref{eq:auxB} will favor head-on
collisions between interacting particles. {At the boundary of the
  cluster, such torques lower the tendency of particles to rotate and
  leave the cluster.} It will play little role in the gas phase, where
there are few collisions. For the particles inside the dense arrested
clusters, $B_i$ is also small by symmetry (see
Fig.~\ref{fig:symmetry}). Therefore, we expect that the dynamics with
the control torque \eqref{eq:auxB} will lead to a reinforcement of
MIPS and hence a lower active work, with a cost function
$(B_i^\theta)^2/4D$ nearly vanishing outside the boundaries of the
cluster.

Simulations using the control force \eqref{eq:auxB} indeed show
reduced numbers of gas-phase particles in Fig~\ref{fig:clustering}(a,b), leading to smaller values of
the active work when compared to the original
dynamics. Fig~\ref{fig:clustering}(c) shows that, furthermore, as the
system sizes are increased, the upper bound of the LDF $\Phi(w)$
strongly decreases. These numerical results support our theory because
they illustrate how a phase-separated arrested state can indeed be
stabilised using a cost that is dominated by boundary
contributions. Note that for much larger sizes, however, our cost
function might saturate because the torque $B_i^{\theta}$ does not
vanish exactly in the bulk of the cluster and gas phases. Only a
protocol that would be exactly restricted to the boundary region
could be used to achieve the $N\to\infty$ limit, which is anyway far
beyond what we can do numerically.

\bibliographystyle{plain_url}
\bibliography{draft_g.bib}

\end{document}